\documentclass[superscriptaddress,aps,prd,showpacs]{revtex4}

\usepackage{amsfonts}
\usepackage{makeidx}
\usepackage{amsmath}
\usepackage{array}
\usepackage{amssymb}
\usepackage{latexsym}
\usepackage{graphicx}
\usepackage{color}
\usepackage{soul}
\usepackage{arydshln}
\usepackage{multirow}
\usepackage[rightcaption]{sidecap}
\usepackage{float}
\usepackage{hyperref}
\usepackage{cleveref}

\begin{document}

\setstcolor{black}

\title{Electromagnetic description of three-dimensional time-reversal invariant ponderable topological insulators}
\author{A. Mart\'{i}n-Ruiz}
\email{alberto.martin@nucleares.unam.mx}
\affiliation{Instituto de Ciencias Nucleares, Universidad Nacional Aut\'{o}noma de M\'{e}xico, 04510 M\'{e}xico, Distrito Federal, M\'{e}xico}

\author{M. Cambiaso}
\affiliation{Universidad Andres Bello, Departamento de Ciencias Fisicas, Facultad de
Ciencias Exactas, Avenida Republica 220, Santiago, Chile}

\author{L. F. Urrutia}
\affiliation{Instituto de Ciencias Nucleares, Universidad Nacional Aut\'{o}noma de M\'{e}%
xico, 04510 M\'{e}xico, Distrito Federal, M\'{e}xico}

\date{\today}

\begin{abstract}
A general technique to analyze the classical interaction between ideal topological insulators, and electromagnetic sources and fields, has been previously elaborated. Nevertheless it is not immediately applicable in the laboratory as it fails to describe real ponderable media. In this work we provide a description of real topologically insulating materials taking into account their dielectric and magnetic properties. For inhomogeneous permittivity and permeability, the problem of finding the Green's function must be solved in an ad hoc manner. Nevertheless, the physically feasible cases of piecewise constant $\varepsilon, \mu$ and $\theta$ make the problem tractable, where $\theta$ encodes the topological magnetoelectric polarizability properties of the medium. To this end we employ the Green's function method to find the  fields resulting form the interaction between these materials and electromagnetic sources. Furthermore we exploit the fact that in the cases here studied, the full Green's function can be successfully found if the  Green's function of the corresponding ponderable media with $\theta = 0$ is known. Our results,  satisfactorily reproduce previously existing ones and also generalize some others.  The method 
here elaborated can be exploited to determine the electromagnetic fields for more general configurations  aiming to measure the  interaction between real 3D topological insulators and electromagnetic fields.

\end{abstract}


\maketitle

\section{Introduction}
\label{introduction}

Most states of quantum matter are described by the symmetries they break \cite{Anderson}. However, topological states of quantum matter evade traditional symmetry-breaking classification schemes. Instead they are  described in the low-energy limit by topological field theories (TFT) \cite{Qi-ReviewTI, Hasan-ReviewTI}. Recently, topological insulators (TIs) have attracted great attention in condensed matter physics \cite{Wang, Maciejko-PRL, Zhou}. These materials, among other unique electronic properties,  display nontrivial topological order and are characterized by a fully insulating bulk and gapless edge or surface states, which are protected by time-reversal symmetry.

In addition to their interesting electronic properties, TIs also exhibit interesting properties in terms of their interaction with electromagnetic sources and fields, in contrast to ordinary insulators or conductors. Most of these properties are a consequence of the topological magnetoelectric effect (TME), which consists in the transmutation of the electric and magnetic induction fields, even in the case of static and stationary sources. The electromagnetic response of a conventional insulator is governed by Maxwell's equations derived from the ordinary electromagnetic Lagrangian $\mathcal{L} _{0} = (1 / 8 \pi) [ \varepsilon \textbf{E} ^{2} - (1 / \mu ) \textbf{B} ^{2} ]$. Three-dimensional TIs are well described by adding a term of the form $\mathcal{L} _{\theta} = (\alpha / 4 \pi ^{2}) \theta \textbf{E} \cdot \textbf{B}$, where $\alpha$ is the fine-structure constant and $\theta$ is the topological magnetoelectric polarizability (TMEP) \cite{Qi-PRB, Essin}. 

Specific TMEs that would result from this ($3+1$)-dimensional TFT have been predicted and they include: induced mirror magnetic monopole-like fields due to electric charges close to the surface of a TI (and vice versa) \cite{Qi-Science, Rosenberg}, a nontrivial Faraday rotation of the polarizations of electromagnetic waves propagating through a TIs surface \cite{Hehl, Maciejko, Huerta}  and a tunable Casimir repulsion between three-dimensional TIs \cite{Grushin-PRL, Rodriguez-PRL}. However, none of these effects has yet been observed experimentally. %
A possible explanation for this is that in $3+1$ dimensions the $\theta$ term and the Maxwell term are equally important at low energies. Another possible explanation is that so far most experimental setups devised to measure these effects do not necessarily favor an eventual $\theta$ signal as fairly simple configurations have been considered. 

To this end, in Refs.~\cite{MCU1, MCU3} we initiated a method to calculate the electromagnetic response of planar, spherical and cylindrical ideal TIs under electromagnetic fields  by means of  Green's functions (GF). In those works, previously existing results have been recovered by generalizing the methods employed in the literature. That generalized description, however, was lacking a more thorough and realistic assessment of TIs since we analyzed the idealized situation in which TIs had neither dielectric nor permeable properties. From a practical point of view, the aforementioned results are instructive but not directly applicable. Therefore the corresponding description of topologically insulating ponderable media is needed. In this work we aim to fill in this gap. Following Refs.~\cite{MCU1, MCU3}  we extend the  GF method to study the TME of time-reversal (TR) invariant TIs including its dielectric and permeable properties, for planar and spherical geometries.

The work is structured as follows. In \cref{3D-TIs} we briefly review the basics of the electromagnetic response of 3-dimensional TR invariant TIs. 
\Cref{Sec_PlanarTI} is devoted to derive the GF for a planar TI occupying the semi-infinite region $z < 0$ in contact with the vacuum, following closely the Ref.~\cite{Schwinger}. This allows us to deal  not only with a piecewise constant TME polarizability as we did in Refs.~\cite{MCU1} but also with piecewise constant electric polarizability and magnetic permeability.  The corresponding differential equations for the reduced GF fall into three groups. We solve the first group to illustrate the procedure, and the solutions for the remaining groups are left for \cref{Appendix}. We will also derive the GF in coordinate representation. In  \cref{Sec_spherTI} a similar calculation is done for the case of a spherical topological insulator.
\Cref{Applications} is dedicated to different applications, the problems of a pointlike electric charge,  an infinitely straight current-carrying wire and  an infinitely straight uniformly charged wire all three configurations  in front of an infinitely planar TI. As a fourth application we then solve the pointlike charge in front of a spherical TI. Finally, in \cref{discussion} we emphasize the contribution of this work in regards its applicability in real situations and elaborate on some general features of the method and its results. We also  discuss the physical  interpretation of the GF in terms  of the image electric and magnetic charge and current densities. Throughout the paper, Lorentz-Heaviside units are assumed ($\hbar =c=1$), the metric signature will be taken as $\left( + , - , - , - \right) $ and the convention $\epsilon ^{0123}=+1$ is adopted.

\section{Electromagnetic response of 3D topological insulators}

\label{3D-TIs}

The electromagnetic response of a conventional insulator is characterized by the dielectric permittivity $\varepsilon$ and the magnetic permeability $\mu$. An electric field induces an electric polarization, whereas a magnetic field induces a magnetic polarization. As both the electric field $\textbf{E} = - \nabla \phi - \frac{\partial \textbf{A}}{\partial t} $ and the magnetic induction $\textbf{B} = \nabla \times \textbf{A}$ are well defined inside the insulator, the linear response of a conventional insulator in the presence of external electric charge $\rho$ and current $\textbf{J}$ densities is determined by the usual action  \cite{Schwinger}  from which the field equations and constitutive relations are derived. 
To describe the interaction between a 3D topologically insulating ponderable media with electromagnetic sources and fields, the  latter must be supplemented by the magneto-electric contribution \cite{Qi-PRB, Essin}.
Thus the action under consideration in this paper is
\begin{equation}
S  = S _{0} + S _{\theta}  = \int d ^{4} x \left( \mathcal{L} _{0} +\mathcal{L} _{\theta}  - \rho \phi + \textbf{J} \cdot \textbf{A}  \right) = \int d ^{4} x \left[ \frac{1}{8 \pi} \left( \varepsilon \textbf{E} ^{2} - \frac{1}{\mu} \textbf{B} ^{2} \right) + \frac{\alpha }{4 \pi ^{2} } \theta \textbf{E} \cdot \textbf{B}- \rho \phi + \textbf{J} \cdot \textbf{A} \right] . \label{FullAction}
\end{equation}
where $\alpha = e ^{2} / \hbar c$ is the fine structure constant and $\theta$ is the TMEP. Under periodic boundary conditions, the partition function and all physical quantities are invariant under shifts of $\theta$ by any multiple of $2 \pi$. Since $\textbf{E} \cdot \textbf{B}$ is odd under TR, there are only two values of $\theta$ which give a TR symmetric theory, namely, $\theta = 0$ and $\theta = \pi$ (mod $2 \pi$) \cite{Karch}. One can therefore conclude that there are two different classes of TR invariant TIs in 3D, the topologically trivial class with $\theta = 0$ and the topologically nontrivial class with $\theta = \pi$. In this work we will consider the nontrivial case only.

The  Maxwell's equations including the topological term  are the usual ones 
\begin{align}
\nabla \cdot \textbf{D} = 4 \pi \rho \phantom{0} \quad & , \quad \nabla \times \textbf{H} = \frac{\partial \textbf{D}}{\partial t} + 4 \pi \textbf{J} , \label{MaxwellEquations} \\ \nabla \cdot \textbf{B} = 0 \phantom{4 \pi \rho} \quad & , \quad \nabla \times \textbf{E} = - \frac{\partial \textbf{B}}{\partial t} , 
\end{align}
with,  however, modified constitutive relations, since $\theta$ is understood as an effective parameter much as $\varepsilon$ and $\mu$:

\begin{equation}
\mathbf{D} = 4 \pi \frac{\delta \mathcal{L}}{\delta \textbf{E}} = \varepsilon \mathbf{E} + \frac{\alpha}{\pi} \theta \mathbf{B} \quad , \quad  \mathbf{H} = - 4 \pi \frac{\delta \mathcal{L}}{\delta \textbf{B}} = \frac{\mathbf{B}}{\mu} - \frac{\alpha}{\pi} \theta \mathbf{E} . \label{ConstitutiveRelationsTI}
\end{equation}

\begin{figure}[h]
\begin{center}
\includegraphics
{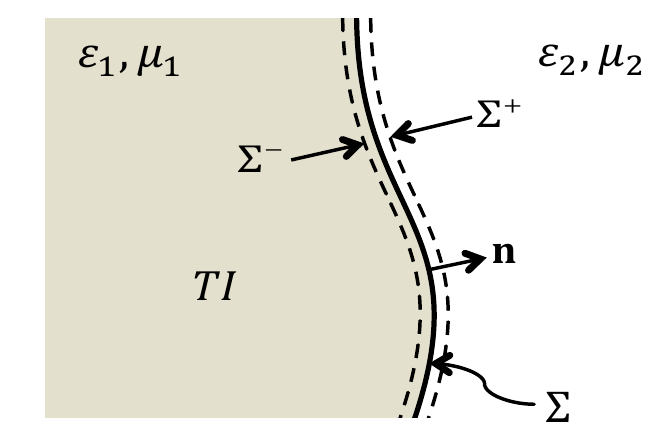}
\end{center}
\caption{{\protect\small Geometry of semi-infinite time-reversal symmetric topological insulator ($\theta \neq 0$) in contact with the vacuum ($\theta = 0$). }}
\label{FIG-planar_TI}
\end{figure}

Assuming that the time derivatives of the fields are finite in the vicinity of the surface $\Sigma$ of a nontrivial 3D TI in contact with a trivial insulator (or vacuum), the  Maxwell's equations imply boundary conditions, which, for vanishing external sources on $\Sigma$, read
\begin{align}
\left[ \varepsilon \mathbf{E} \right] _{\Sigma} \cdot \textbf{n} = \tilde{\theta} (\mathbf{B} \cdot \textbf{n} ) \big| _{\Sigma} \quad &, \quad \left[ \tfrac{1}{\mu} \mathbf{B} \right] _{\Sigma} \times \textbf{n} = - \tilde{ \theta} ( \mathbf{E} \times \textbf{n} ) \big| _{\Sigma} ,  \label{Ampere-BC} \\  \left[ \mathbf{B} \right] _{\Sigma} \cdot \textbf{n} = 0 \;\;\;\;\;\;\;\;\;\;\;\;\;\ \quad &, \quad \;\;\;\; \left[ \mathbf{E} \right] _{\Sigma} \times \textbf{n} = 0 .  \label{Faraday-BC}
\end{align}
These are derived by integrating the field equations over a pill-shaped region across $\Sigma$, with $\tilde{\theta} = \alpha \theta / \pi$. The notation is $\left[ \textbf{V} \right] _{\Sigma} = \textbf{V} (\Sigma ^{+}) - \textbf{V} (\Sigma ^{-})$, where $\textbf{n}$ is the unit normal to $\Sigma$ and the surfaces $\Sigma ^{+}$, $\Sigma ^{-}$ are all defined in Fig. \ref{FIG-planar_TI}. 

The continuity conditions  Eqs.~(\ref{Faraday-BC}) imply that the right hand sides of Eqs.~(\ref{Ampere-BC}) are well defined and they represent self-induced surface charge and current densities, respectively. An immediate consequence of these boundary conditions is the transmutations of the electric and magnetic induction fields, which characterizes the  TME of TR invariant 3D TIs.

\section{Green's function method}
\label{GF-method}

In this section we adapt the Green's function method initiated in Refs.~\cite{MCU1} to study the TME of TR invariant 3D TIs endowed with dielectric and permeable properties. 
Knowledge of the GF allows one to compute the electromagnetic fields for an arbitrary distribution of sources, as well as to solve problems with given Dirichlet, Neumann or Robin boundary conditions on arbitrary surfaces. The advantages of the GF method are apparent in the case of ordinary electrodynamics of continuous media and so is the case for the electromagnetics of TIs. For example, in Refs.~\cite{MCU1} some potential benefits from the experimental standpoint have been mentioned. Also, in Ref.~\cite{MCU3}, conceptual and computational virtues of the method were analyzed in the context of the Casimir attraction between a TI and a conducting plate. For the method to be applicable in a real experimental situation, though, the dielectric and permeable properties of the TIs must be taken into account. This implies a generalization of the method elaborated in Refs.~\cite{MCU1}. The subtleties arise as one needs to find the GF for a position dependent $\varepsilon(\textbf{r})$ and $\mu(\textbf{r})$, which we will take as piecewise constant functions, in analogy with the TMEP $\theta$. This will be one of the main results of this work. 

Since the homogeneous Maxwell's equations that relate the potentials to the fields are not modified in the presence of the $\theta$ term, the electrostatic and magnetostatic fields can be written in terms of the electromagnetic 4-potential $A ^{\mu} = (\phi , \textbf{A})$ according to $\textbf{E} = - \nabla \phi$ and $\textbf{B} = \nabla \times \textbf{A}$, as usual. In the Coulomb gauge $\nabla \cdot \textbf{A} = 0$, the electromagnetic potentials satisfy the coupled equations of motion
\begin{align}
- \varepsilon (\textbf{r}) \nabla ^{2} \phi - \nabla \varepsilon (\textbf{r}) \cdot \nabla \phi + \frac{\alpha}{\pi} \nabla \theta (\textbf{r}) \cdot \nabla \times \textbf{A} &= 4 \pi \rho , \label{MaxEqs-Potentials} \\ - \tilde{\mu} (\textbf{r}) \nabla ^{2} \textbf{A} + \frac{\alpha}{\pi} \nabla \theta (\textbf{r}) \times \nabla \phi + \nabla \tilde{\mu} (\textbf{r}) \times \left( \nabla \times \textbf{A} \right) &= 4 \pi \textbf{J} .  
\label{MaxEqs-Potentials2}
\end{align}
where $\tilde{\mu} (\textbf{r}) = 1 / \mu (\textbf{r})$. If the TMEP is homogeneous, $\theta (\textbf{r}) = \theta$, the equations of motion for the scalar $\phi$ and vector $\textbf{A}$ potentials  decouple (since $\nabla \theta (\textbf{r}) = 0$), and one finds that the propagation of the electromagnetic fields is the same as in a conventional insulator. As a consequence, electromagnetic waves propagating within a 3D TI retain their usual properties: dispersion is linear, the phase and group velocities are proportional to the refractive index, the fields are transverse and orthogonal polarizations do not mix \cite{Maciejko}. Therefore, the effects of the $\theta$ term can be measured only when the TMEP varies in space.

\subsection{Planar topological insulator}
\label{Sec_PlanarTI}

Let us restrict ourselves to the analysis of a planar TI. Without loss of generality we choose our coordinates such that $\textbf{n} = \hat{\textbf{e}} _{z}$, where $\hat{\textbf{e}} _{z}$ is the outward unit normal to $\Sigma$ which is located at $z=0$. This represents the case of a homogeneous TI occupying the region $z < 0$ in contact with the vacuum ($z > 0$). Then the inhomogeneity in $\theta (\textbf{r})$ is limited to a finite discontinuity across the surface $\Sigma$, such that $\nabla \theta (\textbf{r}) = - \theta \delta(z) \hat{\textbf{e}} _{z} $. Given the above choice, it is also natural to take the dependence upon position of the ponderable properties of the media as $\varepsilon (\textbf{r}) = \varepsilon H (- z) + H (z)$ and $\tilde{\mu} (\textbf{r}) = \frac{1}{\mu} H (- z) + H (z) $ respectively. Therefore, the field equations describing the TME of a planar TR symmetric 3D TI can be written in the matrix form
\begin{equation}
\left[ \mathcal{O} ^{\mu} _{\phantom{\mu} \nu} \right] _{\textbf{x}}  A ^{\nu} (\textbf{x} ) = 4 \pi J ^{\mu} , \label{MaxEqs-Potentials-Matrix}
\end{equation}
where $J ^{\mu} = (\rho , \textbf{J})$ is a conserved external 4-current, and the differential operator $\left[ \mathcal{O} ^{\mu} _{\phantom{\mu} \nu} \right] _{\textbf{x}}$ reads directly from the equations of motion (\ref{MaxEqs-Potentials})-(\ref{MaxEqs-Potentials2}), \textit{i.e.}

\begin{align}
\left[ \mathcal{O} ^{\mu} _{\phantom{\mu} \nu} \right] _{\textbf{x}} =  \left[ \begin{array}{cccc} \mathcal{O} ^{(\varepsilon)}  & \tilde \theta \delta(z) \frac{\partial}{\partial y} & - \tilde \theta \delta(z) \frac{\partial}{\partial x} & 0 \\  \tilde \theta \delta(z) \frac{\partial}{\partial y} & \mathcal{O} ^{(\tilde{\mu})} & 0 & \left( 1 - \tilde{\mu} \right) \delta (z) \frac{\partial}{\partial x} \\ - \tilde \theta \delta(z) \frac{\partial}{\partial x} & 0 & \mathcal{O} ^{(\tilde{\mu})} & \left( 1 - \tilde{\mu} \right) \delta (z) \frac{\partial}{\partial y} \\ 0 & 0 & 0 & - \tilde{\mu} (z) \nabla ^{2} \end{array} \right] , \label{O-operator}
\end{align}
where 
\begin{align}
\mathcal{O} ^{(f)} = - \nabla \cdot \left[ f (z) \nabla \right] = - f (z) \nabla ^{2} - \frac{\partial f (z)}{\partial z} \frac{\partial}{\partial z}
\end{align}
and $f$ is either the permittivity $\varepsilon$ or permeability $\tilde \mu$.

To obtain a general solution for the 4-potential $A ^{\mu} (\textbf{x})$ in the presence of arbitrary external sources $J ^{\mu} (\textbf{x})$, we introduce the Green's function $G ^{\mu} _{\phantom{\mu} \nu} (\textbf{x} , \textbf{x} ^{\prime})$ solving Eq.~(\ref{MaxEqs-Potentials-Matrix}) for a pointlike source,
\begin{equation}
\left[ \mathcal{O} ^{\mu} _{\phantom{\mu} \nu} \right] _{\textbf{x}}  G ^{\nu} _{\phantom{\nu} \sigma} (\textbf{x} , \textbf{x} ^{\prime} ) = 4 \pi \eta ^{\mu} _{\phantom{\mu} \sigma} \delta (\textbf{x} - \textbf{x} ^{\prime} ) , \label{TI-GF/Matrix-Eq}
\end{equation}
in such a way that the general solution for the 4-potential in the Coulomb gauge is
\begin{equation}
A ^{\mu} (\textbf{x}) = \int G ^{\mu} _{\phantom{\mu} \nu} (\textbf{x} , \textbf{x} ^{\prime}) J ^{\nu} ( \textbf{x} ^{\prime} ) d ^{3} \textbf{x} ^{\prime} . \label{A-GF}
\end{equation}
In the following, we discuss the general solution to Eq.~(\ref{TI-GF/Matrix-Eq}). The GF we consider has translational invariance in the directions parallel to $\Sigma$, that is in the transverse $x$ and $y$ directions, while this invariance is broken in the $z$ direction. Exploiting this symmetry we further introduce the Fourier transform in the direction parallel to the plane $\Sigma$, taking the coordinate dependence to be $\textbf{R} = ( x -x ^{\prime} , y - y ^{\prime})$ and define 
\begin{equation}
G ^{\mu} _{\phantom{\mu} \nu} \left( \mathbf{x} , \mathbf{x} ^{\prime }\right) = 4 \pi \int \frac{d ^{2} \mathbf{p}}{\left( 2 \pi \right) ^{2}}e ^{i \mathbf{p} \cdot \textbf{R}} g ^{\mu} _{\phantom{\mu} \nu} \left( z , z ^{\prime }\right) ,  \label{RedGFDef}
\end{equation}
where $\mathbf{p} = ( p _{x} , p _{y} )$ is the momentum parallel to $\Sigma $ \cite{Schwinger}. In Eq.~(\ref{RedGFDef}) we have omitted the dependence of the reduced GF $g ^{\mu} _{\phantom{\mu} \nu}$ on $\mathbf{p}$. The reduced GF introduced in the above representation satisfies the equation
\begin{equation}
\left[ \mathcal{O} ^{\mu} _{\phantom{\mu} \nu} \right] _{z} g ^{\nu} _{\phantom{\nu} \sigma} \left( z , z ^{\prime }\right) = \eta ^{\mu} _{\phantom{\mu} \sigma} \delta (z - z ^{\prime}) , \label{RedGF-Eq}
\end{equation} 
where $\left[ \mathcal{O} ^{\mu} _{\phantom{\mu} \nu} \right] _{z}$ is the differential operator defined in Eq.~(\ref{O-operator}) with the replacements $\partial _{x} \rightarrow i p _{x}$, $\partial _{y} \rightarrow i p _{y}$ and $- \nabla ^{2} \rightarrow \Box ^{2} \equiv \textbf{p} ^{2} - \partial ^{2} _{z}$. In this way $\mathcal{O} ^{(f)} \rightarrow \mathcal{O} ^{(f)} = f (z) \Box ^{2} - \frac{\partial f (z)}{\partial z} \frac{\partial}{\partial z}$. We now must solve the reduced GF equation for the various components along the same lines introduced in Refs.~\cite{MCU1, MCU3}. 
We observe that the corresponding equations for the reduced GF fall into three groups. The first is defined by $\sigma = 0$, the second by $\sigma = i$, with $i = 1,2$, and the third with $\sigma = 3$ in Eq.~(\ref{RedGF-Eq}). The four equations of the first group (the solution for the remaining groups is left for \cref{Appendix}) read:
\begin{align}
& \mathcal{O} ^{(\varepsilon)} g ^{0} _{\phantom{0} 0} - i \tilde{\theta} \delta (z) \epsilon ^{0i \phantom{j} 3} _{\phantom{0i} j} p _{i} g ^{j} _{\phantom{j} 0} = \delta ( z - z ^{\prime} ) , \label{G1/G00} \\[5pt] & \mathcal{O} ^{(\tilde{\mu})} g ^{i} _{\phantom{i} 0} - i \tilde{\theta} \delta (z) \epsilon ^{0ij3} p _{j} g ^{0} _{\phantom{0} 0} = 0 , \label{G1/Gi0}
\end{align}
where $i,j=1,2$. Note that we take $g ^{3} _{\phantom{3}0} = 0$, as suggested by the solution of the third group ($\sigma = 3$) of equations together with the symmetry property 
\begin{align}
G _{\mu \nu} (\textbf{x} , \textbf{x} ^{\prime}) = G _{\nu \mu} (\textbf{x} ^{\prime} , \textbf{x}) . \label{reciprocity}
\end{align}
Certainly the case of inhomogeneous ponderable media is interesting in its own \cite{Dzyaloshinskii2}. These inhomogeneities are given by the form of the functions $\varepsilon (z)$ and $\mu(z)$. For example, inhomogeneous permittivities with different profiles have been employed for the study of Casimir physics  in Refs.~\cite{Casimir}. In the context of TIs, to the best of our knowledge, mostly homogeneous permittivities are considered.  It is noteworthy that in Ref.~\cite{Crosse:2015loa} a relevant aspect of inhomogeneous permittivities has been mentioned. We will restrict to inhomogeneities in the permittivity, the permeability and $\theta$ only, as other kinds of inhomogeneities e.g., in  temperature, composition, etc, might lead to additional, albeit small, effects that are beyond the scope of this work. Among them is the fact that $\textbf{D}$ may not vanish when $\textbf{E}$ does \cite{LL_EM}. Dispersive effects will also be omitted  from the discussion as only the spatial variation is relevant here. For an arbitrary dependence on $z$ of the permittivity and permeability the solutions to the free reduced GFs of Eqs.~(\ref{G1/G00}) and (\ref{G1/Gi0}) are given in terms of the expected integral equations, which are of no use to write down here. 

Regardless of the specific spatial dependence of $\varepsilon(z)$ and $\mu(z)$, one of  the important lessons from Refs.~\cite{MCU1} is that since  the equation for the reduced GF is linear in $\theta$ and it enters as $\partial_z \theta (z) =-\theta \delta (z)$, then the free reduced GF can be used to integrate the complete GF equation, where by free reduced GF we understand the Green's function corresponding to the operators $\mathcal{O} ^{(\varepsilon)}$ and $\mathcal{O} ^{(\tilde \mu)} $ in Eqs.~(\ref{G1/G00}) and (\ref{G1/Gi0}) respectively, i.e. to the case $\theta = 0$. In other words, if the free reduced Green's functions are given then we can solve for the complete GF of the problem following the same techniques  elaborated in Refs.~\cite{MCU1}. This is precisely the approach that we will use. 

To solve Eqs.~(\ref{G1/G00}) and (\ref{G1/Gi0}) we employ a similar method to that used for obtaining the GF for the one-dimensional $\delta$-function potential in quantum mechanics, where the free GF is used for integrating the GF equation with the $\delta$-interaction. To proceed, we use the free reduced GF $\mathfrak{g} ^{(f)} (z , z ^{\prime})$, associated with the operator $\mathcal{O} ^{(f)}$, that solves
\begin{align}
\mathcal{O} ^{(f)} \mathfrak{g} ^{(f)} (z , z ^{\prime}) = \delta (z - z ^{\prime}) .
\end{align}
The corresponding free reduced GFs of Eqs.~(\ref{G1/G00}) and (\ref{G1/Gi0}) are known \cite{Schwinger}.
\begin{align}
\label{FreeReducedGF}
\begin{split}
\mathfrak{g} ^{(f)} (z , z ^{\prime} > 0) &= \mathfrak{g} ( z , z ^{\prime} ) - \frac{f - 1}{f + 1} \mathfrak{g} ( z , 0) e ^{- p z ^{\prime}} \\ \mathfrak{g} ^{(f)} (z , z ^{\prime} < 0) &= \frac{1}{f} \mathfrak{g} ( z , z ^{\prime} ) + \frac{1}{f} \frac{f - 1}{f + 1} \mathfrak{g} ( z , 0) e ^{p z ^{\prime}} \end{split} \quad  , \quad f = \varepsilon,\, \tilde \mu.
\end{align}
where $p = \vert \textbf{p} \vert $ and $\mathfrak{g} ( z , z ^{\prime} ) = \frac{1}{2p} e ^{- p \vert z - z ^{\prime} \vert}$ is the reduced GF in free space ($\theta = 0$ and $\varepsilon = \mu = 1$) satisfying the standard boundary conditions at infinity. This guarantees  Eqs.~(\ref{FreeReducedGF})   satisfy the corresponding boundary conditions at the vacuum-medium interface.

As mentioned above, we may now exploit the fact that Eqs.~(\ref{G1/G00}) and (\ref{G1/Gi0}) can be directly integrated by using the reduced GF for dielectric and magnetic materials given in Eqs.~(\ref{FreeReducedGF}) together with the properties of the Dirac delta function, thus reducing the problem to a set of coupled algebraic equations
\begin{align}
g ^{0} _{\phantom{0} 0} (z , z ^{\prime}) &= \mathfrak{g} ^{(\varepsilon)} (z , z ^{\prime}) + i \tilde{\theta} \epsilon ^{0i \phantom{j} 3} _{\phantom{0i} j} p _{i} \mathfrak{g} ^{(\varepsilon)} (z , 0) g ^{j} _{\phantom{j} 0} (0 , z ^{\prime}) , \label{g00/int} \\ g ^{i} _{\phantom{i} 0} (z , z ^{\prime}) &= + i \tilde{\theta} \epsilon ^{0ij3} p _{j} \mathfrak{g} ^{(\tilde{\mu})} (z , 0) g ^{0} _{\phantom{0} 0} ( 0 , z ^{\prime}) . \label{gi0/int}
\end{align}

To solve for the various components we set $z = 0$ in Eq.~(\ref{gi0/int}) and then substitute into Eq.~(\ref{g00/int}), yielding
\begin{align}
g ^{0} _{\phantom{0} 0} (z , z ^{\prime}) &= \mathfrak{g} ^{(\varepsilon)} (z , z ^{\prime}) - \tilde{\theta} ^{2} \textbf{p} ^{2}  \mathfrak{g} ^{(\varepsilon)} (z , 0)  \mathfrak{g} ^{(\tilde{\mu})} (0 , 0) g ^{0} _{\phantom{0} 0} ( 0 , z ^{\prime}) , \label{g00/int2}
\end{align}
where we have used the result $\textbf{p} ^{2} = \epsilon ^{0i \phantom{j} 3} _{\phantom{0i} j} \epsilon ^{0jk3} p _{k} p _{i}$. Solving for $g ^{0} _{\phantom{0} 0} (0 , z ^{\prime})$ by setting $z = 0$ in Eq.~(\ref{g00/int2}) and inserting the result back into that equation, we obtain
\begin{align}
g ^{0} _{\phantom{0} 0} (z , z ^{\prime}) &= \mathfrak{g} ^{(\varepsilon)} (z , z ^{\prime}) - \frac{\tilde{\theta} ^{2} \textbf{p} ^{2} \mathfrak{g} ^{(\tilde{\mu})} (0 , 0)}{1 + \tilde{\theta} ^{2} \textbf{p} ^{2}  \mathfrak{g} ^{(\varepsilon)} (0 , 0)  \mathfrak{g} ^{(\tilde{\mu})} (0 , 0) } \mathfrak{g} ^{(\varepsilon)} (z , 0)  \mathfrak{g} ^{(\varepsilon)} (0 , z ^{\prime}) . \label{g00/FIN}
\end{align}
The remaining components can be obtained by substituting $g ^{0} _{\phantom{0} 0} (0 , z ^{\prime})$ in Eq.~(\ref{gi0/int}). The result is
\begin{align}
g ^{i} _{\phantom{i} 0} (z , z ^{\prime}) &= + \frac{i \tilde{\theta} \epsilon ^{0ij3} p _{j}}{1 + \tilde{\theta} ^{2} \textbf{p} ^{2}  \mathfrak{g} ^{(\varepsilon)} (0 , 0)  \mathfrak{g} ^{(\tilde{\mu})} (0 , 0) } \mathfrak{g} ^{(\tilde{\mu})} (z , 0) \mathfrak{g} ^{(\varepsilon)} (0 , z ^{\prime}) . \label{gi0/FIN}
\end{align}
Our results of Eqs.~(\ref{g00/FIN}) and (\ref{gi0/FIN}) can be written in terms of the reduced GF in free space $\mathfrak{g} (z , z ^{\prime})$ as follows:
\begin{align}
g ^{0} _{\phantom{0} 0} (z , z ^{\prime}) &= \frac{1}{\varepsilon (z ^{\prime})} \left[ \mathfrak{g} (z , z ^{\prime}) - \frac{\mbox{sgn} (z ^{\prime}) (\varepsilon - 1) (\frac{1}{\mu} + 1) + \tilde{\theta} ^{2}}{(\varepsilon + 1) (\frac{1}{\mu} + 1) + \tilde{\theta} ^{2}} 4 \textbf{p} ^{2} \mathfrak{g} (0 , 0) \mathfrak{g} (z , 0) \mathfrak{g} (0 , z ^{\prime}) \right] , \label{g00/FIN2} \\ g ^{i} _{\phantom{i} 0} (z , z ^{\prime}) &= + \frac{4 i \tilde{\theta} \epsilon ^{0ij3} p _{j}}{(\varepsilon + 1) (\frac{1}{\mu} + 1) + \tilde{\theta} ^{2}} \mathfrak{g} (z , 0) \mathfrak{g} (0 , z ^{\prime}) . \label{gi0/FIN2}
\end{align}
Using the same procedure one can further solve for the groups defined by $\sigma = j$ (with $j=1,2$) and $\sigma = 3$ in Eq.~(\ref{RedGF-Eq}). The detailed calculations are presented in \cref{Appendix}. The solution is
\begin{align}
g ^{i} _{\phantom{i} j} ( z , z ^{\prime} ) &= \eta ^{i} _{\phantom{i} j} \mu (z ^{\prime}) \left[ \mathfrak{g} (z , z ^{\prime}) - 4 \textbf{p} ^{2} \mathfrak{g}(0,0) \mathfrak{g}(z,0) \mathfrak{g}(0,z ^{\prime}) \frac{\mbox{sgn}(z ^{\prime}) (\varepsilon + 1)(\frac{1}{\mu} - 1) + \tilde{\theta}^{2} }{(\varepsilon + 1) (\frac{1}{\mu} + 1) + \tilde{\theta} ^{2}} \right] \notag \\ & \hspace{2cm} - \left[ \frac{2 \tilde{\theta} ^{2} (\frac{1}{\mu} + 1) ^{-1} }{(\varepsilon + 1) (\frac{1}{\mu} + 1) + \tilde{\theta} ^{2}} + \left( \frac{1 - \mu}{1 + \mu} \right) ^{2} \mu (z ^{\prime})  \right] 4 p ^{i} p _{j} \mathfrak{g}(0,0) \mathfrak{g}(z,0) \mathfrak{g}(0,z ^{\prime})  , \label{G2-gij-FIN} \\ g ^{i} _{\phantom{i} 3} (z , z ^{\prime}) &= \mu (z ^{\prime})  \left[ \eta ^{i} _{\phantom{i} 3} \mathfrak{g}(z , z ^{\prime}) + 2i \frac{1 - \mu}{1 + \mu} p ^{i} \mathfrak{g}(z , 0) \mathfrak{g}(0 , z ^{\prime}) \right] , \label{gi3/FIN}
\end{align}
together with symmetry property $g ^{0} _{\phantom{0} i} ( z , z ^{\prime} ) = g ^{i} _{\phantom{i} 0} ( z , z ^{\prime} )$.

The reciprocity between the position of the unit charge and the position at which the GF is evaluated, Eq.~(\ref{reciprocity}), is one of its most remarkable properties. From Eq.~(\ref{RedGFDef}) this conditions demands
\begin{align} \label{symmredGF}
g _{\mu \nu} (z , z ^{\prime} , \textbf{p}) = g _{\nu \mu} (z ^{\prime} , z , - \textbf{p}) ,
\end{align}
which we verify directly from Eqs.~(\ref{g00/FIN2})-(\ref{gi3/FIN}). The symmetry $g _{\mu \nu} (z , z ^{\prime} , \textbf{p}) = g _{\mu \nu } ^{\ast} (z , z ^{\prime} , - \textbf{p}) $ is a consequence of the reality of the Green's function in Eq.~(\ref{RedGFDef}).

The various components of the static GF matrix in coordinate representation are obtained by computing the Fourier transform defined in Eq. (\ref{RedGFDef}), with the reduced GF given by Eqs. (\ref{g00/FIN2})-(\ref{gi3/FIN}). The required integrals can be evaluated in a simple fashion. The details are presented in Refs.~\cite{MCU1}. Here we remind the reader of the most important results, 

\begin{align}
I ^{0} (\textbf{x} , \textbf{x} ^{\prime}) & = 4 \pi \int \frac{d ^{2} \textbf{p}}{(2 \pi) ^{2}} e ^{i \textbf{p} \cdot \textbf{R}} 4 \textbf{p} ^{2} \mathfrak{g} (0,0) \mathfrak{g} (z,0) \mathfrak{g} (0,z ^{\prime}) = \frac{1}{\sqrt{R ^{2} + Z ^{2}}} , \\ \textbf{I} (\textbf{x} , \textbf{x} ^{\prime}) & = 4 \pi \int \frac{d ^{2} \textbf{p}}{(2 \pi) ^{2}} e ^{i \textbf{p} \cdot \textbf{R}} 4 \textbf{p} \mathfrak{g} (z,0) \mathfrak{g} (0,z ^{\prime}) = 2i \frac{\textbf{R}}{R ^{2}} \left( 1 - \frac{Z}{\sqrt{R ^{2} + Z ^{2}}} \right) , \\ \textbf{K} (\textbf{x} , \textbf{x} ^{\prime}) & = 4 \pi \int \frac{d ^{2} \textbf{p}}{(2 \pi) ^{2}} e ^{i \textbf{p} \cdot \textbf{R}} 8 \textbf{p} \mathfrak{g} (0,0) \mathfrak{g} (z,0) \mathfrak{g} (0,z ^{\prime}) = 2i \frac{\textbf{R}}{R ^{2}} \left( \sqrt{R ^{2} + Z ^{2}} - Z \right) , 
\end{align}
where $Z = \vert z \vert + \vert z ^{\prime} \vert$, $\textbf{R} = (x-x^{\prime} , y-y^{\prime})$ and $R = \vert \textbf{R} \vert$. The static GF in coordinate representation can be written in terms of the previous integrals as
\begin{align}
& G ^{0} _{\phantom{0} 0} (\textbf{x} , \textbf{x} ^{\prime}) =\frac{1}{\varepsilon (z ^{\prime})} \left[ \frac{1}{\vert \textbf{x} - \textbf{x} ^{\prime} \vert} - \frac{\mbox{sgn} (z ^{\prime}) (\varepsilon - 1) (\frac{1}{\mu} + 1) + \tilde{\theta} ^{2}}{(\varepsilon + 1) (\frac{1}{\mu} + 1) + \tilde{\theta} ^{2}} \frac{1}{\sqrt{R ^{2} + Z ^{2}}} \right] , \label{TI-G00} \\ & G ^{i} _{\phantom{i} 0} (\textbf{x} , \textbf{x} ^{\prime}) = \frac{i \tilde{\theta}}{(\varepsilon + 1) (\frac{1}{\mu} + 1) + \tilde{\theta} ^{2}} \epsilon ^{0ij3} I _{j} (\textbf{x}, \textbf{x} ^{\prime}) = G ^{0} _{\phantom{0} i} (\textbf{x} , \textbf{x} ^{\prime}) , \label{TI-Gi0} \\ & G ^{i} _{\phantom{i} j} (\textbf{x} , \textbf{x} ^{\prime}) = \eta ^{i} _{\phantom{i} j} \mu (z ^{\prime}) \left[ \frac{1}{\vert \textbf{x} - \textbf{x} ^{\prime} \vert} - \frac{\mbox{sgn} (z ^{\prime}) (\varepsilon + 1) (\frac{1}{\mu} - 1) + \tilde{\theta} ^{2}}{(\varepsilon + 1) (\frac{1}{\mu} + 1) + \tilde{\theta} ^{2}} \frac{1}{\sqrt{R ^{2} + Z ^{2}}} \right]  \notag \\ & \hspace{3cm} - \frac{i}{2} \left[ \frac{2 \tilde{\theta} ^{2} (\frac{1}{\mu} + 1) ^{-1} }{(\varepsilon + 1) (\frac{1}{\mu} + 1) + \tilde{\theta} ^{2}} + \left( \frac{1 - \mu}{1 + \mu} \right) ^{2} \mu (z ^{\prime}) \right] \partial _{j} K ^{i} (\textbf{x} , \textbf{x} ^{\prime}) , \label{TI-Gij} \\ & G ^{i} _{\phantom{i} 3} (\textbf{x} , \textbf{x} ^{\prime}) = \mu (z ^{\prime}) \left[ \eta  ^{i} _{\phantom{i} 3} \frac{1}{\vert \textbf{x} - \textbf{x} ^{\prime} \vert} + \frac{i}{2} \frac{1 - \mu}{1 + \mu} I ^{i} ( \textbf{x} , \textbf{x} ^{\prime}) \right] . \label{TI-Gi3}
\end{align}
We observe that Eqs.~(\ref{TI-G00})-(\ref{TI-Gi3}) contain all the required elements of the GF matrix, according to the choices of $z$ and $z ^{\prime}$ in the function $Z$. From the experimental point of view, it is interesting the case in which the sources are in vacuum ($z ^{\prime} > 0$); however, our solutions also describe sources embedded in the TI ($z ^{\prime} < 0$). Finally, we can further check the consistency of our results with the previously reported results. Firstly, in the limiting case $\varepsilon = 1$ and $\mu = 1$ these results reduce correctly to the ones reported in Refs.~\cite{MCU1} for a planar $\theta$ boundary; and secondly, if we take $\theta = 0$ the GF becomes the one reported by Schwinger et al. in Ref.~\cite{Schwinger} for a semi-infinite dielectric media.

Previously we mentioned the symmetry $G _{\mu \nu} (\textbf{x} , \textbf{x} ^{\prime}) = G _{\nu \mu} (\textbf{x} ^{\prime} , \textbf{x})$. The particular solutions in  Eqs.~(\ref{TI-G00}) through (\ref{TI-Gi3}), although not manifestly symmetric,  can easily be verified to satisfy this symmetry.

\subsection{Spherical topological insulator}
\label{Sec_spherTI}

Now let us consider the case of a nonmagnetic spherical topological insulator. In this section, following the same procedure as in the planar situation, we discuss the spherical case, in which the values of $\theta$ and $\varepsilon$ have a discontinuity across the surface $r = a$. In the adapted spherical coordinates $(r, \vartheta , \varphi)$, it proves to be convenient to introduce explicitly the angular momentum operator $\hat{\textbf{L}} = \frac{1}{i} \textbf{x} \times \nabla$. In fact, the GF equation can be written as
\begin{equation}
\left[ \mathcal{O} ^{\mu} _{\phantom{\mu} \nu} \right] _{\textbf{x}}  G ^{\nu} _{\; \sigma} (\textbf{x} , \textbf{x} ^{\prime} ) = 4 \pi \eta ^{\mu} _{\phantom{\mu} \sigma} \delta (\textbf{x} - \textbf{x} ^{\prime} ) , \label{GF-Eq-Spherical}
\end{equation}
where the differential operator is given by Eq. (\ref{TI-O-operator-Esfericas}). Since the square of momentum angular operator commutes with this operator, its solution has the form
\begin{equation}
G ^{\mu} _{\phantom{\mu} \nu} (\textbf{x} , \textbf{x} ^{\prime} ) = 4 \pi \sum _{l=0} ^{\infty} \sum _{m=-l} ^{+l} \sum _{m ^{\prime}=-l} ^{+l} g ^{\mu} _{l m m ^{\prime} , \nu} (r , r ^{\prime} ) Y _{lm} (\vartheta , \varphi) Y _{lm ^{\prime}} ^{\ast} (\vartheta ^{\prime} , \varphi ^{\prime}) ,  \label{TI-GF-Esfericas-text}
\end{equation}
with the reduced GF $g ^{\mu} _{l m m ^{\prime} , \nu} (r , r ^{\prime} )$ satisfying the equation 
\begin{align}
\sum _{m ^{\prime \prime}=-l} ^{+l} \hat{R} ^{\mu} _{l m m ^{\prime \prime} , \nu}  g ^{\nu} _{l m ^{\prime \prime} m ^{\prime} , \sigma} (r , r ^{\prime} ) = \eta ^{\mu} _{\phantom{\mu}  \sigma} \delta _{m m ^{\prime}} \frac{\delta (r - r ^{\prime})}{r ^{2}} , \label{TI-GF/Matrix-Eq-Esfericas3}
\end{align}
where $\hat{R} ^{\mu} _{ l m m ^{\prime \prime} , \nu} = \langle l m \vert \left[ \mathcal{O} ^{\mu} _{\phantom{\mu} \nu} \right] _{\textbf{x}} \vert l m ^{\prime \prime} \rangle$. This equation can be integrated in the same way  as for the planar symmetry. The detailed calculation is presented in appendix \ref{AppSpherical}. The solution for the various components is
\begin{align}
g ^{0} _{l m  m ^{\prime} , 0} (r , r ^{\prime} ) &= \delta _{mm ^{\prime}} \left[ \mathfrak{g} ^{(\varepsilon)} _{l} (r , r ^{\prime} ) - a ^{2} \tilde{\theta} ^{2} l (l+1) \mathfrak{g} ^{(1)} _{l} (a , a ) S _{l} ^{( \varepsilon , \varepsilon)} (r , r ^{\prime}) \right] , \label{g00-esf-F} \\ g ^{i} _{l m  m ^{\prime} , 0} (r , r ^{\prime} ) &= - i a \tilde{\theta}  \langle \hat{\textbf{L}} ^{i} \rangle _{m m ^{\prime}} S _{l} ^{( 1 , \varepsilon)} (r , r ^{\prime}) , \label{gi0-esf-F} \\ g ^{0} _{l m  m ^{\prime} , i} (r , r ^{\prime} ) &= + i a \tilde{\theta} \langle \hat{\textbf{L}} _{i} \rangle _{m m ^{\prime}} S _{l} ^{( \varepsilon , 1)} (r , r ^{\prime}) , \label{g0i-esf-F} \\ g ^{i} _{l m  m ^{\prime} , j} (r , r ^{\prime} ) &= \eta ^{i} _{\phantom{i} j} \delta _{mm ^{\prime}} \mathfrak{g} ^{(1)} _{l} (r , r ^{\prime} ) + a ^{2} \tilde{\theta} ^{2} \langle \hat{\textbf{L}} ^{i} \hat{\textbf{L}} _{j} \rangle _{mm ^{\prime}} \mathfrak{g} ^{(\varepsilon)} _{l} (a , a ) S _{l} ^{( 1,1)} (r , r ^{\prime}) , \label{gij-esf-F}
\end{align}
where
\begin{align}
S _{l} ^{( f , g)} (r , r ^{\prime}) = \frac{\mathfrak{g} _{l} ^{(f)} (r,a) \mathfrak{g} _{l} ^{(g)} (a,r ^{\prime})}{1 + a ^{2} \tilde{\theta}^2 l (l+1) \mathfrak{g} _{l} ^{(1)} (a,a) \mathfrak{g} _{l} ^{(\varepsilon)} (a,a)} \quad , \quad f,g=1, \varepsilon . \label{S-function}
\end{align}
Here $\mathfrak{g} _{l} ^{(\varepsilon)} (r,r ^{\prime})$ is the reduced GF in the absence of the $\theta$-term given by Eqs.~(\ref{gel_1})-(\ref{gel_4}) and $\mathfrak{g} _{l} ^{(1)} (r,r ^{\prime})$ is the reduced GF in free space given in Eq.~(\ref{g1l_freeSpher}).

The reciprocity symmetry $G _{\mu \nu} (\textbf{x} , \textbf{x} ^{\prime}) = G _{\nu \mu} (\textbf{x} ^{\prime} , \textbf{x})$, together with the reality of the GF, $G _{\mu \nu} (\textbf{x} , \textbf{x} ^{\prime}) = G ^{\ast} _{\mu \nu} (\textbf{x}  , \textbf{x} ^{\prime})$, imply the following conditions upon the reduced GF:
\begin{align}
g _{l m  m ^{\prime} , \mu \nu} (r , r ^{\prime} ) = (-1) ^{m + m ^{\prime}} g _{l -m ^{\prime} -m, \nu \mu} (r ^{\prime} ,r) \quad , \quad g _{l m  m ^{\prime} , \mu \nu} (r , r ^{\prime} ) = (-1) ^{m + m ^{\prime}} g ^{\ast} _{l -m - m ^{\prime}, \mu \nu} (r ,r ^{\prime}) ,
\end{align}
respectively. These results can be combined to give the generalized hermiticity condition
\begin{align}
g _{l m  m ^{\prime} , \mu \nu} (r , r ^{\prime} ) = g ^{\ast} _{l m ^{\prime} m, \nu \mu} (r ^{\prime} ,r) . \label{SYM}
\end{align}
One can further verify that the explicit expressions (\ref{g00-esf-F})-(\ref{gij-esf-F}) satisfy the relation (\ref{SYM}).

\section{Applications} \label{Applications}

\subsection{Pointlike charge near a planar TI}

\begin{figure}[h]
\begin{center}
\includegraphics{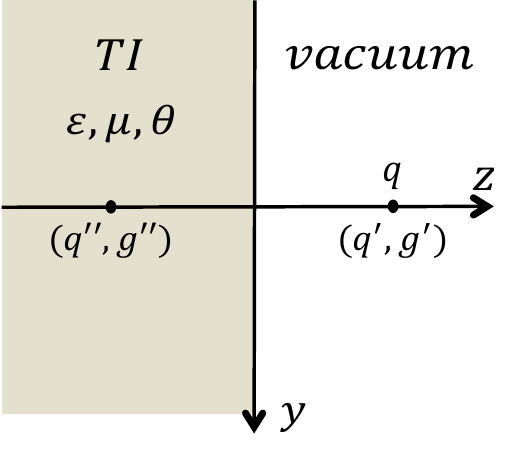}\qquad
\end{center}
\caption{{\protect\small Illustration of the images electric charges and magnetic monopoles induced by a pointlike electric charge near to a planar topological insulator.}}
\label{Carga-TIPlano}
\end{figure}

Let us consider the geometry as shown in Fig.~\ref{Carga-TIPlano}. The left-half space ($z<0$) is occupied by a TR invariant TI with dielectric constant $\varepsilon$,  magnetic permeability $\mu$ and  TMEP $\theta$, whereas the right-half space ($z>0$) is the vacuum. A pointlike electric charge $q$ is located in vacuum at a distance $b > 0$ from the TI. We can always choose the coordinates such that $x ^{\prime} = y ^{\prime} = 0$, in such a way that the current density is $j ^{\mu} \left( \mathbf{x}^{\prime }\right) = q \eta ^{\mu} _{\phantom{\mu} 0} \delta \left( x ^{\prime} \right) \delta \left( y ^{\prime} \right) \delta \left( z ^{\prime} - b \right) 
$. 

According to Eq.~(\ref{A-GF}), the solution for this problem is
\begin{equation}
A ^{\mu} \left( \mathbf{x} \right) = q G ^{\mu} _{\phantom{\mu} 0} \left( \mathbf{x} , \mathbf{r} \right) ,  \label{TI-SolPointCharge}
\end{equation}
where $\mathbf{r} = b \hat{\mathbf{e}} _{z}$. We first study the electrostatic potential. From Eq.~(\ref{TI-G00}), we obtain the $00$-component of the GF,
\begin{align}
z > 0 \quad & : \quad G ^{0} _{\phantom{0} 0} (\textbf{x} , \textbf{r}) = \frac{1}{\vert \textbf{x} - \textbf{r} \vert} - \frac{(\varepsilon - 1) (\frac{1}{\mu} + 1) + \tilde{\theta} ^{2}}{(\varepsilon + 1) (\frac{1}{\mu} + 1) + \tilde{\theta} ^{2}} \frac{1}{\vert \textbf{x} + \textbf{r} \vert} ,  \label{TI-G00z>} \\ z < 0 \quad & : \quad G ^{0} _{\phantom{0} 0} (\textbf{x} , \textbf{r}) = \frac{2 (\frac{1}{\mu} + 1)}{(\varepsilon + 1) (\frac{1}{\mu} + 1) + \tilde{\theta} ^{2}} \frac{1}{ \vert  \mathbf{x} - \mathbf{r} \vert } .
\label{TI-G00z<}
\end{align}
For $z>0$ the GF yields  the electric potential  $A ^{0} \left( \mathbf{x} \right) = q G ^{0} _{\phantom{0} 0} \left( \mathbf{x}, \mathbf{r}
\right)$ which can be interpreted as due to two pointlike electric charges, one of strength $q$ at $\mathbf{r}$, and the other, the image charge, of strength 
\begin{equation}
q ^{\prime \prime} = - q \frac{(\varepsilon - 1) (\frac{1}{\mu} + 1) + \tilde{\theta} ^{2}}{(\varepsilon + 1) (\frac{1}{\mu} + 1) + \tilde{\theta} ^{2}} ,
\end{equation}
at the point $- \mathbf{r}$. For $z<0$ only one pointlike electric charge appears, of strength 
\begin{equation}
q ^{\prime} = q + q ^{\prime \prime} = \frac{2 q (\frac{1}{\mu} + 1)}{(\varepsilon + 1) (\frac{1}{\mu} + 1) + \tilde{\theta} ^{2}} , 
\end{equation}
located at $\mathbf{r}$.

From Eq.~(\ref{TI-SolPointCharge}) we see that two components of the magnetic vector potential are nonzero, $A ^{1} \left( \mathbf{x} \right)
= q G ^{1} _{\phantom{1} 0} \left( \mathbf{x} , \mathbf{r} \right)$ and $A ^{2} \left( \mathbf{x}\right) = q G ^{2} _{\phantom{2} 0} \left( \mathbf{x} , \mathbf{r} \right)$. According to Eq.~(\ref{TI-Gi0}) the corresponding GF components for each region are given by
\begin{align}
G ^{1} _{\phantom{1} 0} \left( \mathbf{x},\mathbf{r} \right) &= + \frac{ 2 \tilde{\theta}}{(\varepsilon + 1)(\frac{1}{\mu} + 1) +\tilde{\theta}^{2} } \frac{y}{R^{2}} \left[ 1 - \frac{\vert z \vert + b}{\vert \textbf{x} + \mbox{sgn}(z) \textbf{r} \vert } \right] , \label{TI-G01-App} \\ G ^{2} _{\phantom{2} 0} \left( \mathbf{x},\mathbf{r} \right) &= - \frac{ 2 \tilde{\theta}}{(\varepsilon + 1)(\frac{1}{\mu} + 1) +\tilde{\theta}^{2} } \frac{x}{R^{2}} \left[ 1 - \frac{\vert z \vert + b}{\vert \textbf{x} + \mbox{sgn}(z) \textbf{r} \vert } \right] , \label{TI-G02-App}
\end{align}
from which we can compute  the magnetic field, $\mathbf{B}=\nabla \times \mathbf{A}$, to  obtain:
\begin{align}
z > 0 \quad &: \quad \mathbf{B} \left( \mathbf{x}\right) = - \frac{ 2 q \tilde{\theta}}{(\varepsilon + 1)(\frac{1}{\mu} + 1) + \tilde{\theta}^{2} } \frac{\mathbf{x}+\mathbf{r}}{ \vert \mathbf{x}+\mathbf{r} \vert ^{3}} , \\ z < 0 \quad &: \quad \mathbf{B} \left( \mathbf{x}\right) = + \frac{ 2 q \tilde{\theta}}{(\varepsilon + 1)(\frac{1}{\mu} + 1) +\tilde{\theta}^{2} } \frac{\mathbf{x}-\mathbf{r}}{\vert \mathbf{x} -\mathbf{r} \vert ^{3}}.
\end{align}
Thus, we observe that the magnetic field for $z>0$ can be interpreted as that of a magnetic monopole of strength 
\begin{equation}
g ^{\prime \prime} = - \frac{ 2 q \tilde{\theta}}{(\varepsilon + 1)(\frac{1}{\mu} + 1) +\tilde{\theta}^{2} }
\end{equation}
located at $- \mathbf{r}$. Similarly, for $z<0$ the magnetic field can be understood as originated by a monopole of strength $g ^{\prime} = - g ^{\prime \prime}$ located at $\mathbf{r}$. These interpretations and the apparent contradiction with $\nabla \cdot \textbf{B} = 0$ will be commented in \cref{discussion}. Our results are in a agreement with the ones reported in Ref.~\cite{Qi-Science}, where the image method was used.

Now let us compute the force between the electric charge and the TI by two different methods, using the interaction energy calculated via the GF  and the energy-momentum tensor approach. The interaction energy between a charge-current distribution and a topological insulator is 
\begin{equation}
E _{int} = \frac{1}{2} \int d \mathbf{x} \int d \mathbf{x} ^{\prime} j ^{\mu} \left( 
\mathbf{x} \right) \left[ G _{\mu \nu} \left( \mathbf{x} ,\mathbf{x} ^{\prime} \right) - \eta _{\mu \nu} \mathcal{G} \left( \mathbf{x} , \mathbf{x} ^{\prime} \right) \right] j ^{\nu} \left( \mathbf{x} ^{\prime} \right) ,
\end{equation}
where $\mathcal{G} \left( \mathbf{x} , \mathbf{x}^{\prime} \right) = 1 / \vert \mathbf{x} - \mathbf{x} ^{\prime} \vert $ is the GF in
vacuum \cite{MCU1}. The first contribution represents the total energy of a charge-current distribution in the presence of the TI, including mutual interactions. We evaluate this energy for the problem of a pointlike electric charge at $\mathbf{r} = b \hat{\mathbf{e}} _{z}$. Making use of Eq.~(\ref{TI-G00z>}), the interaction energy is
\begin{equation}
E _{int} = - \frac{q ^{2}}{4b} \frac{(\varepsilon - 1) (\frac{1}{\mu} + 1) + \tilde{\theta} ^{2}}{(\varepsilon + 1) (\frac{1}{\mu} + 1) + \tilde{\theta} ^{2}} . 
\end{equation}
Our result implies that the force on the charge exerted by the TI is
\begin{equation}
\textbf{F} = - \frac{\partial E _{int}}{\partial b} \hat{\textbf{e}} _{z} = - \frac{q ^{2}}{(2b) ^{2}} \frac{(\varepsilon - 1) (\frac{1}{\mu} + 1) + \tilde{\theta} ^{2}}{(\varepsilon + 1) (\frac{1}{\mu} + 1) + \tilde{\theta} ^{2}} \hat{\textbf{e}} _{z} , \label{force1}
\end{equation}
noting that it is always attractive. This  can be interpreted as the force between the charge $q$ and the image charge $q ^{\prime \prime}$ according to Coulomb's law. The field theory point of view provides an alternative derivation by computing the net flux of momentum over a close surface $S$ enclosing the charge. We take $S$ as $\Sigma ^{+}$ (just outside the TI, at $z = 0 ^{+}$) plus a semi-sphere at infinity, where the electromagnetic fields are zero. In terms of the stress-energy tensor this force is
\begin{equation}
\mathbf{F} = - \hat{\mathbf{e}} _{z} \int _{\Sigma ^{+}} d S T _{zz} \left( \Sigma
^{+} \right) .  \label{INTST}
\end{equation}
As discussed in Refs.~\cite{MCU1, MCU3}, the stress tensor has the form as that in standard electrodynamics, but as expected, it is not conserved on the surface of the TI because the self-induced charge and current densities arising there. Thus, the required expression for $T_{zz}\left( \Sigma ^{+}\right)$ in Eq.~(\ref{INTST}) is the standard one 
\begin{equation}
T _{zz} = \frac{1}{8 \pi} \left[ E _{\parallel} ^{2} - E _{z} ^{2} + B _{\parallel} ^{2} - B _{z} ^{2}\right] ,
\end{equation}
where $E _{z}$ ($B _{z}$) denotes the electric (magnetic) field component normal to the surface and $E _{\parallel}$ ($B _{\parallel}$) is the component of the electric (magnetic) field parallel to the surface. According to our results, the electric and magnetic fields for $z > 0$ are
\begin{align}
\mathbf{E} \left( \mathbf{x}\right) &= q \frac{\mathbf{x} - \mathbf{r}}{\vert \mathbf{x} - \mathbf{r} \vert ^{3}} - q \frac{(\varepsilon - 1) (\frac{1}{\mu} + 1) + \tilde{\theta} ^{2}}{(\varepsilon + 1) (\frac{1}{\mu} + 1) + \tilde{\theta} ^{2}} \frac{\mathbf{x} + \mathbf{r}}{\vert \mathbf{x} + \mathbf{r} \vert ^{3}},  \label{Efield} \\ \mathbf{B}\left( \mathbf{x} \right) &= - \frac{ 2 q \tilde{\theta}}{(\varepsilon + 1)(\frac{1}{\mu} + 1) + \tilde{\theta} ^{2}} \frac{\mathbf{x} + \mathbf{r}}{\vert \mathbf{x} + \mathbf{r} \vert ^{3}}. \label{Bfield}
\end{align}
Thus we find 
\begin{align}
\mathbf{F} &= \frac{q ^{2}}{\left[ (\varepsilon + 1)(\frac{1}{\mu} + 1) + \tilde{\theta} ^{2} \right] ^{2}} \hat{\mathbf{e}} _{z} \int _{0} ^{\infty} dR\frac{R}{\left( R ^{2} + b ^{2} \right) ^{3}} \left\lbrace R ^{2} \left( \frac{1}{\mu} + 1 \right) ^{2} - b ^{2} \left[ \varepsilon \left( \frac{1}{\mu} + 1 \right) + \tilde{\theta} ^{2} \right] ^{2} + \tilde{\theta} ^{2} \left( R ^{2} - b ^{2} \right) \right\rbrace \notag \\ &= - \frac{q ^{2}}{(2b) ^{2}} \frac{(\varepsilon - 1) (\frac{1}{\mu} + 1) + \tilde{\theta} ^{2}}{(\varepsilon + 1) (\frac{1}{\mu} + 1) + \tilde{\theta} ^{2}} \hat{\textbf{e}} _{z} ,
\end{align}
in agreement with Eq.~(\ref{force1}).

\subsection{Infinitely straight current-carrying wire near a planar TI}

\label{current-wire}

\begin{figure}[h]
\begin{center}
\includegraphics{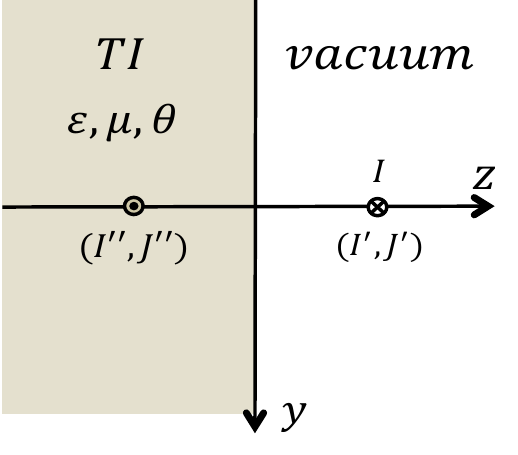} \qquad
\end{center}
\caption{{\protect\small Illustration of the images electric and magnetic current-carrying wires induced by an infinitely straight current-carrying wire near a planar topological insulator.}}
\label{FIG-Current-Wire}
\end{figure}

Now let us consider  an infinitely straight wire parallel to the $x$ axis and carrying a current $I$ in the $+x$ direction. The wire is located in vacuum at a distance $b$ from an semi-infinite TI with a dielectric constant $\varepsilon$, a magnetic permeability $\mu$ and a TMEP $\theta$, as shown in Fig.~\ref{FIG-Current-Wire}. Choosing the coordinates such that $y^{\prime }=0$, the current density is $ j ^{\mu} \left( \mathbf{x} ^{\prime} \right) = I \eta ^{\mu} _{\phantom{\mu} 1} \delta \left( y ^{\prime} \right) \delta \left( z ^{\prime} - b \right)$.

The solution for this problem can be written in terms of the GF as follow
\begin{equation}
A ^{\mu} (\textbf{x}) = I \int _{- \infty} ^{+ \infty} G ^{\mu} _{\phantom{\mu} 1} (\textbf{x} , \textbf{r}) dx ^{\prime} , \label{TI-CurrentWire}
\end{equation}
where $\textbf{r} = x^{\prime} \hat{\textbf{e}} _{x} + b \hat{\textbf{e}} _{z}$. Clearly the nonzero component $A ^{0} (\textbf{x}) = I \int _{- \infty} ^{+ \infty} G ^{0} _{\phantom{0} 1} (\textbf{x} , \textbf{r}) dx ^{\prime}$ arising from the GF implies that an electric field is induced, where $G ^{0} _{\phantom{0}1}$, defined in Eq.~(\ref{TI-Gi0}) is given by
\begin{equation}
G ^{0} _{\phantom{0} 1} (\textbf{x} , \textbf{r}) = + \frac{2 \tilde{\theta}}{(\varepsilon + 1) (\frac{1}{\mu} +1) + \tilde{\theta} ^{2}} \frac{y}{R ^{2}} \left[ 1 - \frac{\vert z \vert + b}{\sqrt{R ^{2} + (\vert z \vert + b) ^{2}}} \right] , \label{01GreenCurrent-TI}
\end{equation}
with $R ^{2} = (x - x ^{\prime}) ^{2} + y ^{2}$. Substituting Eq.~(\ref{01GreenCurrent-TI}) into Eq.~(\ref{TI-CurrentWire}) yields the electric potential, which lacks an immediate interpretation. We can directly compute the electric field as $\textbf{E} (\textbf{x}) = - \nabla A ^{0} (\textbf{x})$, with the result
\begin{equation}
\mathbf{E} \left( \mathbf{x} \right) = - \frac{4 \tilde{\theta}I}{(\varepsilon + 1) (\frac{1}{\mu} +1) + \tilde{\theta} ^{2}} \left[ \frac{\vert z \vert + b}{y ^{2} + \left( \vert z \vert + b \right) ^{2}} \mathbf{\hat{e}} _{y} - \frac{y \; \mbox{sgn} \left( z \right)}{y ^{2} + \left( \vert z \vert + b \right) ^{2}} \mathbf{\hat{e}} _{z} \right] .
\end{equation}
We observe that the electric field for $z > 0$ is equivalent to that due to an infinitely straight wire parallel to the $x$ axis, located inside the TI at $z = - b$, and carrying a magnetic current 
\begin{equation}
J ^{\prime \prime} = \frac{2 \tilde{\theta}}{(\varepsilon + 1) (\frac{1}{\mu} +1) +\tilde{\theta}^{2}} I 
\end{equation}
in the $-x$ direction. For $z < 0$ the field is as if produced by an infinitely straight wire parallel to the $x$ axis, located in vacuum at $z = b$, and carrying a magnetic current $J ^{\prime} = J ^{\prime \prime}$ in the $+x$ direction. Seeming violation of Faraday's law will be commented in \cref{discussion}.

Similarly, now we compute the magnetic field. The nonzero components of the magnetic vector potential are $A ^{1} =  I \int _{- \infty} ^{+ \infty} G ^{1} _{\phantom{1} 1} (\textbf{x} , \textbf{r}) dx ^{\prime}$ and $A ^{2} =  I \int _{- \infty} ^{+ \infty} G ^{2} _{\phantom{2} 1} (\textbf{x} , \textbf{r}) dx ^{\prime}$, where the corresponding GF components are given by
\begin{align}
& G ^{1} _{\phantom{1} 1} (\textbf{x} , \textbf{r}) = \frac{1}{\sqrt{R ^{2} + \vert z - b \vert ^{2}}} - \frac{(\varepsilon + 1) (\frac{1}{\mu} - 1) + \tilde{\theta} ^{2}}{(\varepsilon + 1) (\frac{1}{\mu} + 1) + \tilde{\theta} ^{2}} \frac{1}{\sqrt{R ^{2} + (\vert z \vert + b) ^{2}}} \notag  \\ & \hspace{1cm} + \left[ \frac{2 \tilde{\theta} ^{2} (\frac{1}{\mu} + 1) ^{-1} }{(\varepsilon + 1) (\frac{1}{\mu} + 1) + \tilde{\theta} ^{2}} + \left( \frac{1 - \mu}{1 + \mu} \right) ^{2} \right] \frac{\partial}{\partial x} \left\lbrace \frac{x-x ^{\prime}}{R ^{2}} \left[ \sqrt{R ^{2} + (\vert z \vert + b) ^{2}} - (\vert z \vert + b) \right] \right\rbrace , \\ & G ^{2} _{\phantom{1} 1} (\textbf{x} , \textbf{r}) = \left[ \frac{2 \tilde{\theta} ^{2} (\frac{1}{\mu} + 1) ^{-1} }{(\varepsilon + 1) (\frac{1}{\mu} + 1) + \tilde{\theta} ^{2}} + \left( \frac{1 - \mu}{1 + \mu} \right) ^{2} \right] \frac{\partial}{\partial x} \left\lbrace \frac{y}{R ^{2}} \left[ \sqrt{R ^{2} + (\vert z \vert + b) ^{2}} - (\vert z \vert + b) \right] \right\rbrace ,
\end{align}
where $R ^{2} = (x - x ^{\prime}) ^{2} + y ^{2}$ as before. 
Again, the magnetic field $\textbf{B} (\textbf{x}) = \nabla \times \textbf{A} = - \partial _{z} A ^{2} \hat{\textbf{e}} _{x} + \partial _{z} A ^{1} \hat{\textbf{e}} _{y} + ( \partial _{x} A ^{2} - \partial _{y} A ^{1} ) \hat{\textbf{e}} _{z}$ can be readily computed and interpreted directly in terms of images. In fact
\begin{align}
& \mathbf{B} \left( \mathbf{x} \right) = I \frac{- 2 ( z - b ) \mathbf{\hat{e}} _{y} + 2y \mathbf{\hat{e}} _{z}}{y ^{2} + \left( z - b \right) ^{2}} - I \frac{(\varepsilon + 1) (\frac{1}{\mu} - 1) + \tilde{\theta} ^{2}}{(\varepsilon + 1) (\frac{1}{\mu} + 1) + \tilde{\theta} ^{2}} \frac{- 2 ( \vert z \vert + b ) \mbox{sgn} (z) \mathbf{\hat{e}} _{y} + 2 y \mathbf{\hat{e}} _{z}}{y ^{2} + \left( \vert z \vert + b \right) ^{2}} .
\end{align}
For $z > 0$ the magnetic field corresponds to the one produced by two infinitely straight wires parallel to the $x$ axis, one carrying an electric current $I$ located in vacuum at $z = + b$, and the other, the image current, located inside the TI at $z = - b$, with current
\begin{equation}
I ^{\prime \prime} = \frac{(\varepsilon + 1) (\frac{1}{\mu} - 1) + \tilde{\theta} ^{2}}{(\varepsilon + 1) (\frac{1}{\mu} + 1) + \tilde{\theta} ^{2}} I ,
\end{equation}
which flows in the $-x$ direction. For $z < 0$ the magnetic field is produced by an infinitely straight wire parallel to the $x$ axis, located in vacuum at $+b$, and carrying an electric current 
\begin{equation}
I ^{\prime} = I - I ^{\prime \prime} = \frac{2 (\varepsilon + 1) }{(\varepsilon + 1) (\frac{1}{\mu} + 1) + \tilde{\theta} ^{2}} I,
\end{equation}
flowing in the $+ x$ direction. 

We observe that in the limiting case in which $\varepsilon = \mu = 1$ in the whole space, we recover correctly the electromagnetic fields reported in Refs.~\cite{MCU1} for an infinitely straight current-carrying wire near a planar $\theta$ boundary. On the other hand, in the limit $\theta = 0$, we obtain that $\textbf{E} (\textbf{x}) = 0$ (absence of the TME), and the image electric currents reduce to
\begin{equation}
I ^{\prime \prime} = \frac{1 - \mu}{1 + \mu} I \qquad \mbox{and} \qquad I ^{\prime} = \frac{2 \mu }{1 + \mu} I ,
\end{equation}
as expected.

\subsection{Infinitely uniformly charged wire near a planar TI}

\begin{figure}[h]
\begin{center}
\includegraphics{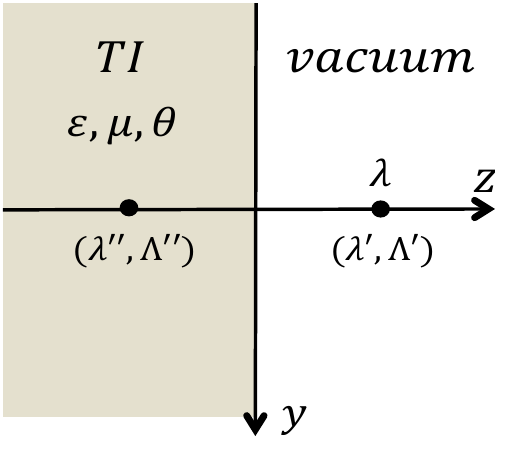} \qquad
\end{center}
\caption{{\protect\small Illustration of the images electric and magnetic charge densities induced by an infinitely uniformly charged wire near a planar topological insulator.}} \label{Corriente-Plano-TI}
\end{figure}

Now we consider an infinite straight wire which carries the uniform charge per unit length $\lambda $. The wire is placed parallel to the $x$-axis and is located in vacuum at a distance $b$ from the TI, as shown in Fig.~\ref{Corriente-Plano-TI}. Choosing the coordinates such that $y ^{\prime}=0$, the current density is $j ^{\mu} \left( \mathbf{x} ^{\prime} \right) = \lambda \eta ^{\mu} _{\phantom{\mu} 0} \delta \left( y ^{\prime} \right) \delta \left( z ^{\prime} - b \right)$. 

The solution for this problem is
\begin{equation}
A ^{\mu} (\textbf{x}) = \lambda \int _{- \infty} ^{+ \infty} G ^{\mu} _{\phantom{\mu} 0} (\textbf{x} , \textbf{r}) dx ^{\prime} , \label{TI-ChargeWire}
\end{equation}
where $\textbf{r} = x ^{\prime} \hat{\textbf{e}} _{x} + b \hat{\textbf{e}} _{z}$. The scalar potential is $A ^{0} (\textbf{x}) = \lambda \int _{- \infty} ^{+ \infty} G ^{0} _{\phantom{0} 0} (\textbf{x} , \textbf{r}) dx ^{\prime}$, where the $00$-component of the GF is given by
\begin{equation}
G ^{0} _{\phantom{0} 0} (\textbf{x} , \textbf{r}) = \frac{1}{\sqrt{R ^{2} + \vert z - b \vert ^{2}}} - \frac{(\varepsilon - 1) (\frac{1}{\mu} + 1) + \tilde{\theta} ^{2}}{(\varepsilon + 1) (\frac{1}{\mu} + 1) + \tilde{\theta} ^{2}} \frac{1}{\sqrt{R ^{2} + (\vert z \vert + b ) ^{2}}} , \label{00GreenCurrent-TI}
\end{equation}
with $R ^{2} = (x - x ^{\prime}) ^{2} + y ^{2}$. The electric field is given by $\textbf{E} (\textbf{x}) = - \nabla A ^{0} (\textbf{x}) =  - \lambda \int _{- \infty} ^{+ \infty} \nabla G ^{0} _{\phantom{0} 0} (\textbf{x} , \textbf{r}) dx ^{\prime} $ resulting in
\begin{align}
\mathbf{E} \left( \mathbf{x} \right) = \lambda \frac{2 y \hat{\textbf{e}} _{y} + 2 (z-b) \hat{\textbf{e}} _{z}}{y ^{2} + (z-b) ^{2}} - \lambda \frac{(\varepsilon - 1) (\frac{1}{\mu} + 1) + \tilde{\theta} ^{2}}{(\varepsilon + 1) (\frac{1}{\mu} + 1) + \tilde{\theta} ^{2}} \frac{2 y \hat{\textbf{e}} _{y} + 2 (\vert z \vert + b) \mbox{sgn} (z) \hat{\textbf{e}} _{z}}{y ^{2} + ( \vert z \vert + b ) ^{2}} .
\end{align}
The electric field for $z>0$ can be interpreted as the one produced by two infinite straight wires, one  with uniform charge per unit length $\lambda$ at $z = + b$, and the other, the image charged wire, with uniform charge per unit length 
\begin{equation}
\lambda ^{\prime \prime} = - \frac{(\varepsilon - 1) (\frac{1}{\mu} + 1) + \tilde{\theta} ^{2}}{(\varepsilon + 1) (\frac{1}{\mu} + 1) + \tilde{\theta} ^{2}} \lambda 
\end{equation}
located inside the TI at $z = - b$. For $z < 0$ the electric field is as if due to an infinite straight wire which carries the uniform charge per unit length
\begin{equation}
\lambda ^{\prime} = \lambda + \lambda ^{\prime \prime} = \frac{2  (\frac{1}{\mu} + 1)}{(\varepsilon + 1) (\frac{1}{\mu} + 1) + \tilde{\theta} ^{2}} \lambda ,
\end{equation}
located at $z = +b$.


Similarly we compute the magnetic field. The nonzero components of the vector potential are $A ^{1} =  \lambda \int _{- \infty} ^{+ \infty} G ^{1} _{\phantom{1} 0} (\textbf{x} , \textbf{r}) dx ^{\prime}$ and $A ^{2} =  \lambda \int _{- \infty} ^{+ \infty} G ^{2} _{\phantom{2} 0} (\textbf{x} , \textbf{r}) dx ^{\prime}$, where the corresponding GF components are given by
\begin{align}
& G ^{1} _{\phantom{1} 0} (\textbf{x} , \textbf{r}) =  \frac{2 \tilde{\theta}}{(\varepsilon + 1) (\frac{1}{\mu} + 1) + \tilde{\theta} ^{2}} \frac{y}{R ^{2}} \left( 1 - \frac{\vert z \vert + b}{\sqrt{R ^{2} + (\vert z \vert + b) ^{2}}} \right) ,  \\ & G ^{2} _{\phantom{2} 0} (\textbf{x} , \textbf{r}) = -  \frac{2 \tilde{\theta}}{(\varepsilon + 1) (\frac{1}{\mu} + 1) + \tilde{\theta} ^{2}} \frac{x - x ^{\prime}}{R ^{2}} \left( 1 - \frac{\vert z \vert + b}{\sqrt{R ^{2} + (\vert z \vert + b) ^{2}}} \right) .
\end{align}
In this case the magnetic field is $\textbf{B} (\textbf{x}) = \nabla \times \textbf{A} = - \partial _{z} A ^{2} \hat{\textbf{e}} _{x} + \partial _{z} A ^{1} \hat{\textbf{e}} _{y} + ( \partial _{x} A ^{2} - \partial _{y} A ^{1} ) \hat{\textbf{e}} _{z}$  and the result is
\begin{align}
& \mathbf{B} \left( \mathbf{x} \right) = - \lambda \frac{2 \tilde{\theta}}{(\varepsilon + 1) (\frac{1}{\mu} + 1) + \tilde{\theta} ^{2}} \frac{2 y \;  \mbox{sgn} (z) \mathbf{\hat{e}} _{y} + 2 ( \vert z \vert + b) \mathbf{\hat{e}} _{z}}{y ^{2} + \left( \vert z \vert + b \right) ^{2}} .
\end{align}
For $z > 0$ the magnetic field would correspond to that generated by an infinite straight wire which carries the uniform magnetic charge per unit length
\begin{equation}
\Lambda ^{\prime \prime} = -\frac{2 \tilde{\theta}}{(\varepsilon + 1) (\frac{1}{\mu} + 1) + \tilde{\theta} ^{2}} \lambda  ,
\end{equation}
located at $z = - b$. Similarly, for $z<0$ the magnetic field is equivalent to the one produced by  an infinite straight wire  carrying a uniform magnetic charge per unit length $\Lambda ^{\prime} = - \Lambda ^{\prime \prime}$, located at $z = + b$. 
The controversial interpretation of the latter fields 
as a consequence of $\nabla \cdot \textbf{B} = \rho_{m}$ is commented in \cref{discussion}.

We observe that in the limiting case in which $\varepsilon = \mu = 1$ in the whole space, we recover correctly the electromagnetic fields reported in Refs.~\cite{MCU1} for an infinitely uniformly charged wire near a planar $\theta$ boundary. On the other hand, in the limit $\theta = 0$, we obtain that $\textbf{B} (\textbf{x}) = 0$ (absence of the TME), and the image electric charge densities reduce to
\begin{equation}
\lambda ^{\prime \prime} = - \frac{\varepsilon - 1}{\varepsilon + 1} \lambda \qquad , \qquad \lambda ^{\prime} = \frac{2}{\varepsilon + 1} \lambda
\end{equation}
as expected.

\subsection{Pointlike charge near a spherical TI}

\begin{figure}[h]
\begin{center}
\includegraphics{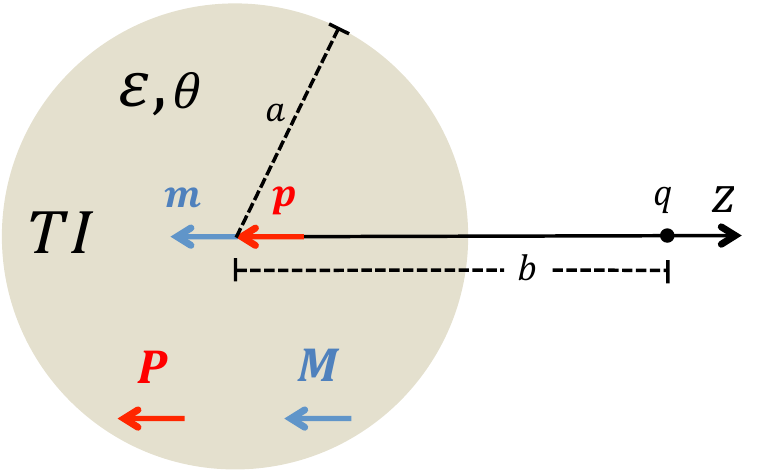} \qquad
\end{center}
\caption{{\protect\small Illustration of the image sources induced by a pointlike electric charge near a spherical TI. In the region $r > b \gg a$ the electric and magnetic fields are due to images electric $\textbf{p}$ and magnetic $\textbf{m}$ dipoles, respectively. Inside the TI ($b \gg a > r$), the electromagnetic fields can be interpreted as those produced by a uniformly polarized sphere with polarization $\mathbf{P}$, while the magnetic field corresponds to the one produced by a uniformly magnetized sphere with magnetization $\mathbf{M}$.}} \label{Esf-Charge-TI}
\end{figure}

The problem we shall discuss is that of a pointlike charge in vacuum located at a distance $b$ from the center of a spherical topological medium of radius $a$, as shown in Fig.~\ref{Esf-Charge-TI}. Choosing the
line connecting the center of the sphere and the charge as the
$z$-axis, the current density can be written as $j ^{\mu} \left(
\mathbf{x} ^{\prime} \right) = \frac{q}{b ^{2}} \eta ^{\mu} _{\; 0} \delta
\left( r ^{\prime} - b \right) \delta \left( \cos \vartheta ^{\prime} - 1
\right) \delta \left( \varphi ^{\prime} \right) $, with $b > a$.

The solution for this problem is then
\begin{align}
\phi \left( \mathbf{x}\right) = q G ^{0} _{\phantom{0} 0} \left(
\mathbf{x} , \mathbf{b} \right) \quad , \quad  \mathbf{A}\left(
\mathbf{x}\right) = q G ^{k} _{\phantom{k} 0} \left( \mathbf{x} ,
\mathbf{b} \right) \hat{\mathbf{e}}_{k},  \label{VectPotSphere}
\end{align}
where $\mathbf{b}=b\hat{\mathbf{e}}_{z}$. With the use of the
corresponding components of the GF matrix, Eqs.~(\ref{g00-esf-F})-(\ref{gij-esf-F}), the scalar and vector potentials become
\begin{align}
\phi \left( \mathbf{x}\right) &= q \sum_{l=0}^{\infty} (2l+1) \left[
\mathfrak{g} _{l} ^{(\varepsilon)} \left( r , b \right) - a ^{2}
\tilde{\theta}^{2} l ( l+1) \mathfrak{g}_{l} ^{(1)} \left( a , a \right) S
_{l} ^{(\varepsilon , \varepsilon)} \left( r , b \right) \right] P_{l}
\left( \cos \vartheta \right) ,  \label{PotSphere3} \\
\mathbf{A}\left( \mathbf{x}\right) &= q \sum_{l=0}^{\infty
}\sum_{m=-l}^{+l} i a \tilde{\theta}\sqrt{4\pi \left( 2l+1\right)}
\left\langle lm\right| \hat{\mathbf{L}}\left| l0\right\rangle S _{l} ^{(1,
\varepsilon)} \left( r , b\right) Y_{lm} \left( \vartheta , \varphi
\right) .  \label{PotVecSphere3}
\end{align}
To derive these results we have used the relations $Y _{lm} \left( 0 ,
\varphi \right) = \delta _{m0}\sqrt{\frac{2l+1}{4 \pi}}$ and $Y _{l0}
\left( \vartheta , \varphi \right) =\sqrt{\frac{2l+1}{4\pi }} P _{l}
\left( \cos \vartheta \right)$. We observe that Eq.~(\ref{PotVecSphere3})
immediately yields $A_{z}=0$, and the remaining components of the vector
potential can be calculated by introducing the combinations $A _{\pm} = A
_{x} \pm i A _{y}$ given by
\begin{equation}
A _{\pm} \left( \mathbf{x} \right) = q \sum _{l=0} ^{\infty} i
a\tilde{\theta}\sqrt{4 \pi ( 2l+1 ) l ( l+1 )} S _{l} ^{(1, \varepsilon)}
\left( r , b \right) Y _{l \pm 1} \left( \vartheta , \varphi \right) ,
\label{PotVecSphere4pm}
\end{equation}
where the expected symmetry $A _{+} = A _{-} ^{\ast}$ follows from the
relation $Y _{l1} ^{\ast} = - Y _{l-1}$. Recalling that
\begin{equation}
A _{\vartheta} = - \sin \vartheta A _{z} + \frac{1}{2} \cos \vartheta
\left( A _{+} e ^{- i \varphi} + A _{-} e ^{i \varphi} \right) \quad ,
\quad A _{\varphi} = \frac{1}{2i} \left( A _{+} e ^{- i \varphi} - A _{-}
e ^{i \varphi } \right) ,
\end{equation}
we obtain
\begin{equation}
A _{\vartheta} = 0 \qquad , \qquad A _{\varphi} = q \sum _{l=0} ^{\infty}
a\tilde{\theta} ( 2l+1) S _{l} ^{(1, \varepsilon)} \left( r , b \right)
\frac{\partial P_{l} \left( \cos \vartheta \right)}{\partial \vartheta },
\label{AFIN}
\end{equation}
in spherical coordinates.

Now we analyze the field strengths for the regions (I) $r>b>a$ and (II) $%
b>a>r$. Using the appropriate reduced Green's functions, $\mathfrak{g}_{l}
^{(1)} \left( r,r^{\prime }\right)$ and $\mathfrak{g}_{l} ^{(\varepsilon)}
\left( r , r ^{\prime} \right)$, the scalar potential and the (nonzero
component of the) vector potential for the region (I) take the form
\begin{align}
\phi _{I}\left( \mathbf{x}\right) &= \frac{q}{\vert \mathbf{x} -\mathbf{b}
\vert} - q \sum _{l=1} ^{\infty} \left\lbrace \frac{(\varepsilon - 1)
l}{(\varepsilon + 1) l + 1} + \frac{\tilde{\theta} ^{2} l (l+1) c
_{l}}{(\varepsilon + 1) l +1} \right\rbrace \frac{a ^{2l+1}}{r ^{l+1} b
^{l+1}} P _{l} \left( \cos \vartheta \right) ,  \label{PotSphere5} \\ A
_{I \varphi}\left( \mathbf{x}\right) &= q \sum_{l=0} ^{\infty}
\tilde{\theta} c _{l} \frac{a ^{2l+1}}{r ^{l+1} b ^{l+1}} \frac{\partial P
_{l} \left( \cos \vartheta \right)}{\partial \vartheta} ,
\label{PotVecSphere5}
\end{align}
respectively. Here
\begin{align}
c _{l} = \frac{2l+1}{(2l+1) [(\varepsilon + 1) l +1] + \tilde{\theta} ^{2}
l ( l+1)} .
\end{align}
The corresponding electric and magnetic fields  can be calculated
directly from $\mathbf{E}=-\nabla \phi $ and $\mathbf{B}=\nabla \times
\mathbf{A}$, respectively. The result is
\begin{align}
\mathbf{E} _{I} \left( \mathbf{x} \right) &= q \frac{\mathbf{x}
-\mathbf{b}}{\vert \mathbf{x} - \mathbf{b} \vert ^{3}} + q \sum _{l =
1}^{\infty} \left\lbrace \frac{(\varepsilon - 1) l}{(\varepsilon + 1) l +
1} + \frac{\tilde{\theta} ^{2} l (l+1) c _{l}}{(\varepsilon + 1) l +1}
\right\rbrace \frac{a ^{2l+1}}{r ^{l+2} b ^{l+1}} \notag \\ & \hspace{3cm}
\left[ - ( l+1) P _{l} \left( \cos \vartheta \right) \hat{\mathbf{r}} +
\frac{\partial P _{l} \left( \cos \vartheta \right)}{\partial \vartheta}
\hat{\boldsymbol{\vartheta}} \right] ,  \label{ElectFieldSphere}
\end{align}
\begin{align}
\mathbf{B}_{I}\left( \mathbf{x}\right) &= q \sum _{l=1} ^{\infty}
\tilde{\theta} l c _{l} \frac{a^{2l+1}}{r^{l+2}b^{l+1}}\left[ -\left(
l+1\right) P_{l}\left( \cos \vartheta \right) \hat{\mathbf{r}}+\frac{%
\partial P_{l}\left( \cos \vartheta \right) }{\partial \vartheta }\hat{\boldsymbol{\vartheta}} \right] .  \label{MagFieldSphere}
\end{align}
We now ask what is the behavior of these field  when the separation
between the point charge and the sphere is large compared to the radius of
the sphere, $b \gg a$. Since the $l$-th term in the sum behaves as $\left(
a/b\right) ^{l+1}$, only small values of $l$ contribute. The leading
contribution in the summations arises from $l=1$ and we obtain
\begin{align}
\mathbf{E} _{I} \left( \mathbf{x}\right) &\sim  q \frac{\mathbf{x}
-\mathbf{b}}{\vert \mathbf{x} - \mathbf{b} \vert ^{3}} + \frac{p}{r
^{3}}\left( 2 \cos \vartheta \hat{\mathbf{r}} + \sin \vartheta
\hat{\boldsymbol{\vartheta}} \right) , \label{ElectFieldSphere2} \\ \mathbf{B}_{I}
\left( \mathbf{x}\right) &\sim \frac{m}{r ^{3}} \left( 2 \cos \vartheta
\hat{\mathbf{r}} + \sin \vartheta \hat{\boldsymbol{\vartheta}} \right) ,
\label{MagFieldSphere2}
\end{align}
which corresponds to the electric and magnetic fields generated by an
electric dipole $\mathbf{p}$ and a magnetic dipole $\mathbf{m}$ lying at
the origin and pointing in the $-z$ direction
\begin{align}
\mathbf{p} &= p \hat{\mathbf{e}}_{z} = - q \left[ \frac{\varepsilon - 1}{\varepsilon + 2} + \frac{6 \tilde{\theta} ^{2}}{\varepsilon + 2} \frac{1}{3(\varepsilon + 2) + 2 \tilde{\theta} ^{2}} \right] \frac{a^{3}}{b^{2}} \hat{\mathbf{e}}_{z} , \\ \mathbf{m} &= m \hat{\mathbf{e}}_{z} = - \frac{3 q \tilde{\theta}}{3(\varepsilon + 2) + 2 \tilde{\theta} ^{2}} \frac{a^{3}}{b^{2}} \hat{\mathbf{e}}_{z} ,
\end{align}
respectively. 

Next we consider the field strengths in the region (II)
$b>a>r$. The scalar
potential and $\varphi $-component of the vector potential become
\begin{align}
\phi _{II} \left( \mathbf{x} \right) &= \frac{q}{\vert
\mathbf{x}-\mathbf{b} \vert} - q \sum _{l=1}^{\infty } \left\lbrace
\frac{(\varepsilon - 1) l}{(\varepsilon + 1) l + 1} + \frac{\tilde{\theta}
^{2} l (l+1) c _{l}}{(\varepsilon + 1) l +1} \right\rbrace
\frac{r^{l}}{b^{l+1}} P _{l} \left( \cos \vartheta \right) ,
\label{PotSphere6} \\ A _{II \varphi }\left( \mathbf{x}\right) &=
q\sum_{l=0}^{\infty }\tilde{\theta} c _{l} \frac{r ^{l}}{b
^{l+1}}\frac{\partial P_{l}\left( \cos \vartheta \right) }{\partial
\vartheta },  \label{PotVecSphere6}
\end{align}
respectively. The corresponding fields are
\begin{align}
\mathbf{E}_{II} \left( \mathbf{x}\right) &= q \frac{\mathbf{x} -
\mathbf{b}}{\vert \mathbf{x} - \mathbf{b} \vert ^{3}} + q \sum
_{l=1}^{\infty } \left\lbrace \frac{(\varepsilon - 1) l}{(\varepsilon + 1)
l + 1} + \frac{\tilde{\theta} ^{2} l (l+1) c _{l}}{(\varepsilon + 1) l +1}
\right\rbrace \frac{r ^{l-1}}{b^{l+1}} \notag \\ & \hspace{3cm} \left[ l P
_{l} \left( \cos \vartheta \right) \hat{\mathbf{r}} + \frac{\partial
P_{l}\left( \cos \vartheta \right) }{\partial \vartheta } \hat{\boldsymbol{\vartheta}}
\right] ,  \label{ElectFieldSphere3} \\ \mathbf{B} _{II} \left( \mathbf{x}
\right) &=- q \sum_{l=1} ^{\infty} \tilde{\theta} (l+1) c _{l}
\frac{r^{l-1}}{b^{l+1}}\left[ l P_{l} \left( \cos \vartheta \right)
\hat{\mathbf{r}} + \frac{\partial P _{l} \left( \cos \vartheta
\right)}{\partial \vartheta } \hat{\boldsymbol{\vartheta}} \right] .
\label{MagFieldSphere3}
\end{align}
When the separation between the point charge and the sphere is large
compared to the radius of the sphere, $b \gg a$, the field strengths in the
region $r<a$ become
\begin{align}
\mathbf{E} _{II} \left( \mathbf{x} \right) & \sim q \frac{\mathbf{x}
-\mathbf{b}}{\vert \mathbf{x} - \mathbf{b} \vert ^{3}} -
\frac{1}{3}\mathbf{P}, \label{ElectFieldSphere4} \\ \mathbf{B} _{II}\left(
\mathbf{x} \right) &\sim \frac{2}{3} \mathbf{M}, \label{MagFieldSphere4}
\end{align}
where
\begin{align}
\mathbf{P} &= - 3 q \left[ \frac{\varepsilon -1}{\varepsilon + 2} +
\frac{6 \tilde{\theta} ^{2}}{3 (\varepsilon + 2) + 2 \tilde{\theta} ^{2}}
\right] \frac{1}{b ^{2}} \hat{\textbf{e}} _{z} , \\ \mathbf{M} &= -
\frac{3}{2} \frac{6 q \tilde{\theta}}{3 (\varepsilon + 2) + 2
\tilde{\theta} ^{2}} \frac{1}{b ^{2}} \hat{\textbf{e}} _{z}
\end{align}
An interesting feature to note is the form of the field  in such
region. The electric field behaves as the field produced by a uniformly
polarized sphere with polarization $\mathbf{P}$, while the magnetic field
resembles the one produced by a uniformly magnetized sphere with
magnetization $\mathbf{M}$.

\section{DISCUSSION}
\label{discussion}

The low-energy effective field theory which describes the electromagnetic response of topological insulators, independently of microscopic details, consists of the usual Maxwell Lagrangian density supplemented by a term proportional to $\theta \textbf{E} \cdot \textbf{B}$, where $\theta$ is the topological magnetoelectric polarizability. Many interesting magnetoelectric effects due to this additional coupling have been highlighted. For example, induced magnetic monopoles due to a pointlike electric charge close to the surface of a planar topological insulator, and a nontrivial Faraday rotation when electromagnetic waves propagates trough such materials. 

Although most of the attention paid to topological insulators is set on their rich electronic structure of quantum mechanical origin, the analysis of their classical interaction with electromagnetic sources and fields is also relevant. For example, it is likely that in a given experiment a topological insulators is made to interact with a macroscopic charged body. In Refs.~\cite{MCU1, MCU3} we initiated a  study along these lines in the idealized case of topological insulators without dielectric and permeable properties. However, to fully account for the electromagnetic response eventual polarizability and magnetization of the topological insulator must be taken into account. In this work we aim to contribute in filling this gap and also to provide concrete applicable methods to find solutions, not presently available in the literature, which could be contrasted with experiments. In particular, for $\varepsilon$, $\mu$  and $\theta$ that vary by a finite discontinuity at the interface between two media, the GF equation can be successfully integrated to yield the complete Green's function of the problem for some specials geometries, in terms of the GF corresponding to the same configuration but with $\theta = 0$. On the other hand, the general advantages of finding the Green's function to determine the electromagnetic fields for a given setup are apparent and can not be underestimated.

In this paper we have constructed the Green's function for planar and spherical 3D time-reversal symmetric topological insulators endowed with ponderable properties and used it to study the electromagnetic fields produced by different simple sources near its surface. Our main results are reported in Eqs.~(\ref{TI-G00})-(\ref{TI-Gi3}) which provide the Green's function for planar symmetry. The corresponding result for the spherical case is given in Eq.~(\ref{TI-GF-Esfericas-text}) together with the reduced Green's functions of Eqs.~(\ref{g00-esf-F})-(\ref{gij-esf-F}).

As a first application, we tackled the problem of a pointlike electric charge near a planar topologically insulating ponderable media, as shown in Fig.~\ref{Carga-TIPlano}. 
We observe that bringing an electric charge into the proximity of  a topological insulator, in addition to the image electric charge an image magnetic monopole will also appear inside the topological insulator. The strengths of the  image electric charges  receive  contributions from the $\theta$ parameter that are additional to the usual ones coming from the optical properties of the medium $\varepsilon$ and $\mu$. On the other hand, the strengths of the image magnetic monopoles depend on $\varepsilon$ and $\mu$ as well, but vanish for $\theta=0$ altogether. This is expected  as there would be no TME effect in that case. As shown in Fig.~\ref{Carga-TIPlano}, the positions of these image electric charges are the same as those in the case of $\theta=0$, in spite of the appearance of the image magnetic monopoles, which are located at the same point as the image electric charges. This is a manifestation of the TME effect which had been reported in Ref.~\cite{Qi-Science}. From this result it follows  that the induced magnetic field has the correct dependence expected from a magnetic monopole, but this solution does not implies the existence of a real magnetic monopole. In fact, one can further check that the magnetic flux integrated over a closed surface that encloses the topological insulator vanishes, consistently with the equation $\nabla \cdot \textbf{B} = 0$. 
The physical origin of the magnetic field is the induced surface current density 
\begin{equation}
\mathbf{J} = \tilde{\theta} \mathbf{E} \times \mathbf{n} \vert _{\Sigma} = - \frac{1}{4 \pi}  \frac{2 q(\frac{1}{\mu} + 1) \tilde{\theta}}{(\varepsilon + 1) (\frac{1}{\mu} + 1) + \tilde{\theta} ^{2}} \frac{R}{\left( R ^{2} + b ^{2} \right) ^{3/2}} \hat{\textbf{e}} _{\varphi} ,
\end{equation}
that is circulating around the origin. It worth mentioning that a similar effect can also occur in other systems with magnetoelectric effect such as multiferroic insulators \cite{Dzyaloshinskii}. However, the topological magnetoelectric effect with a quantized value of $\theta$, and thus the image monopole effect are unique signatures of 3D topological insulators with time-reversal invariance  \cite{Qi-ReviewTI}.

As a second example, we considered the case of an infinitely straight current-carrying wire near a planar topological insulator with parameters $\varepsilon, \mu$ and $\theta$. Similar to the previous case,  the electromagnetic field can also be interpreted in terms of suitable image electric and image magnetic current densities. 
Once again, the results are in good correspondence with the   $\theta=0$ case. Namely, the strengths of the image electric current densities receive $\theta$ contributions additional to the usual ones due to $\varepsilon$ and $\mu$ alone. The strengths of the image magnetic current densities depend on $\varepsilon$ and $\mu$ but are proportional to $\theta$, \textit{i.e.,} there is no TME effect for $\theta = 0$. The positions of images electric and magnetic current densities are shown in Fig. \ref{FIG-Current-Wire}. 

The appearance of a magnetic current in this solution seems to violate the static Faraday's law $\nabla \times \mathbf{E} = 0$, which remained unaltered in the case of the electromagnetic theory extended by the $\theta$ term. However, the electric field results from the induced surface charge density
\begin{equation}
\rho = \tilde{\theta} \mathbf{B} \cdot \mathbf{n} \vert _{\Sigma} =   \frac{1}{\pi}\, \frac{ I \tilde{\theta} (\varepsilon+1)}{(\varepsilon + 1) (\frac{1}{\mu} + 1) + \tilde{\theta} ^{2}}\, \frac{y}{ y^{2} + b ^{2}} .
\end{equation}
The induced electric field has the correct dependence expected from a magnetic current, which is, however, just an artifact to interpret the fields. Indeed, one can directly check that the electric flux vanishes when integrated over a closed surface $S$ that encloses the topological insulator, \textit{i.e.}, $\oint _{S} \textbf{D} \cdot d \textbf{A} = 0$. This result is in agreement with the Gauss's law due to the absence of external charge densities.

We also considered the problem of an infinitely straight uniformly charged wire near a planar 3D topological insulator, where similar interpretations can be made. In this case the electric fields is equivalent to that produced by the original linear charge density plus image lines. The densities of the images have a $\theta$ contribution additional to the usual one and the $\theta=0$ case reproduces the known result.  On the other hand, a magnetic field is induced and interpreted as due to  image magnetic current linear densities. Their strengths depend on $\varepsilon$ and $\mu$ but are proportional to $\theta$. The locations of all image densities are shown in Fig.~\ref{Corriente-Plano-TI}.
It is worth pointing out that this solution does not violate the Maxwell law $\nabla \cdot \textbf{B} = 0$, given that the magnetic field is induced by a surface current density 
\begin{equation}
\textbf{J} = \tilde{\theta} \mathbf{E} \times \mathbf{n} \vert _{\Sigma} = \frac{1}{4 \pi}  \frac{2  \lambda (\frac{1}{\mu} + 1) \tilde{\theta}}{(\varepsilon + 1) (\frac{1}{\mu} + 1) + \tilde{\theta} ^{2}} \frac{2 y}{y ^{2} + b ^{2}} \hat{\textbf{e}} _{x} .
\end{equation}
Such magnetic field can also be interpreted as the one generated by an infinitely uniformly image magnetic charged wire.

The last application is the pointlike charge in vacuum at a distance $b$ from a non-magnetic spherical TI of radius $a$. As in the previous cases the electromagnetic fields are calculated from the GF here obtained. Their closed expressions are given in terms of Legendre polynomials as expected, owing to the spherical geometry with axial symmetry. To facilitate the interpretation of the fields we focus on the cases where $a / b \ll 1$, thus keeping only the dominant contributions in the polynomial expansions. We analyze the behavior of the fields $\textbf{E} (\textbf{r})$ and $\textbf{B} (\textbf{r})$ when $\vert \textbf{r} \vert > b$ and $\vert \textbf{r} \vert < a$, respectively. In the first case, the electric field has the contributions from the charge and from the sphere, which is seen as an electric dipole. The dipole is the usual one of a dielectric sphere plus an additional term depending on $\varepsilon$ and $\theta$. Interestingly the TME effect produces a magnetic field which is as if the charge and the TI sphere induce a magnetic dipole whose magnitude also depends on $\varepsilon$ and $\theta$. Both dipoles are located at the origin and along the symmetry axis. When the field point is inside the sphere the electric field receives an additional contribution to the polarization, and the magnetic field behaves as that of a uniformly magnetized sphere, both arising from the $\theta$-term.

Summarizing, in this work we have provided the general method to determine the electromagnetic fields produced by arbitrary configuration of sources in interaction with real 3D topological insulators for different geometries, as characterized by constant $\varepsilon, \mu$ and $\theta$. Also we can adapt the results of Refs.~\cite{MCU1} to the case of ponderable media, where the electromagnetic fields are sourced by fixed boundary conditions. The  electromagnetic fields 
thus predicted can be used in experimental situations that probe the macroscopic response of topological insulators with the static electromagnetic field.

\acknowledgments 
M.~C. would like to thank the kind hospitality at ICN-UNAM during the preparation of this work. L.~F.~U. and A.~M.~R. have been supported in part by Project No. IN104815 from Direcci\'{o}n General Asuntos del Personal Acad\'{e}mico (Universidad Nacional Aut\'{o}noma de M\'{e}xico) and CONACyT (M\'{e}xico), Project 
No. 237503.

\appendix

\section{GF for a planar TI}
\label{Appendix}
Here we derive Eqs.~(\ref{G2-gij-FIN}) and (\ref{gi3/FIN}) using the same procedure described in \cref{GF-method}. We start with the third group of equations given that it plays a prominent role for the solution of the remaining groups. This group is defined by $\sigma = 3$ in Eq.~(\ref{RedGF-Eq}). The coupled differential equations we have to solve are
\begin{align}
\mathcal{O} ^{(\varepsilon)} g ^{0} _{\phantom{0} 3} + i \tilde{\theta} \delta (z) \epsilon _{0ij3} p ^{i} g ^{j} _{\phantom{j} 3} &= 0 ,  \label{G3-g03} \\ \mathcal{O} ^{(\tilde{\mu})} g ^{i} _{\phantom{i} 3}  - i \tilde{\theta} \delta (z) \epsilon ^{0ij3} p _{j} g ^{0} _{\phantom{0} 3} + i (1 - \tilde{\mu})  p ^{i} \delta (z) g ^{3} _{\phantom{3} 3} &= 0 ,  \label{G3-gi3} \\ \tilde{\mu} (z) \Box ^{2} g ^{3} _{\phantom{3} 3} &= \delta (z - z ^{\prime}) . \label{G3-g33}
\end{align}
These equations can be integrated directly, with the result
\begin{align}
g ^{0} _{\phantom{0} 3} (z,z^{\prime}) &= - i \tilde{\theta} \epsilon _{0ij3} p ^{i} \mathfrak{g} ^{(\varepsilon)} ( z , 0 ) g ^{j} _{\phantom{j} 3} (0,z^{\prime})  ,  \label{G3-gi3/3} \\ g ^{i} _{\phantom{i} 3} (z,z^{\prime}) &=  i \tilde{\theta} \epsilon ^{0ij3} p _{j} \mathfrak{g} ^{(\tilde{\mu})} ( z , 0 ) g ^{0} _{\phantom{0} 3} (0,z^{\prime}) - i (1 - \tilde{\mu}) p ^{i} \mathfrak{g} ^{(\tilde{\mu})} (z,0) g ^{3} _{\phantom{3} 3} (0 , z ^{\prime}) ,  \label{G3-gi3/33} \\ g ^{3} _{\phantom{3} 3} (z , z ^{\prime}) &= \mathfrak{g} (z,z^{\prime}) / \tilde{\mu} (z ^{\prime}) . \label{G3-g33/3}
\end{align}
Substituting Eqs.~(\ref{G3-gi3/3}) and (\ref{G3-g33/3}) into Eq.~(\ref{G3-gi3/33}) yields
\begin{align}
g ^{0} _{\phantom{0} 3} (z,z^{\prime}) = \tilde{\theta} ^{2} \textbf{p} ^{2} \mathfrak{g} ^{(\tilde{\mu})} ( 0 , 0 ) \mathfrak{g} ^{(\varepsilon)} ( z , 0 ) g ^{0} _{\phantom{0} 3} (0,z^{\prime}) ,
\end{align}
from which we conclude that $g ^{0} _{\phantom{0} 3} (z,z^{\prime}) = 0$. Consistency with Eq.~(\ref{G3-gi3/3}) requires $\epsilon _{0ij3} p ^{i} g ^{j} _{\phantom{j} 3} = 0 $. In this way, the solution for the remaining components can be written in terms of the free reduced GF as
\begin{align}
g ^{i} _{\phantom{i} 3} (z,z^{\prime}) = 2i \frac{1 - \mu}{1 + \mu} p ^{i} \mathfrak{g}(z , 0) \mathfrak{g}(0 , z ^{\prime}) / \tilde{\mu} (z ^{\prime})  ,  \label{G3-gi3-FIN}
\end{align}
which clearly satisfies the required condition $\epsilon ^{0j \phantom{k}3} _{\phantom{0j}k}p _{j} g ^{k} _{\phantom{k}3} \propto \epsilon ^{0j \phantom{k}3} _{\phantom{0j}k}p _{j} p ^{k}= 0$.

Now we consider the second group of equations, which is defined by $\sigma = i$ ( with $i=1,2$) in Eq.~(\ref{RedGF-Eq}),
\begin{align}
& \mathcal{O} ^{(\varepsilon)} g ^{0} _{\phantom{0} i} - i \tilde{\theta} \delta (z) \epsilon ^{0j \phantom{k}3} _{\phantom{0j}k} p _{j} g ^{k} _{\phantom{k} i}= 0 , \label{G2-g0i} \\ & \mathcal{O} ^{(\tilde{\mu})} g ^{i} _{\phantom{i} j} - i \tilde{\theta} \delta (z)  \epsilon ^{0ik3} p _{k} g ^{0} _{\phantom{0} j} + i (1 - \tilde{\mu}) p ^{i} \delta (z) g ^{3} _{\phantom{3} j} = \eta ^{i} _{\phantom{i}j} \delta (z - z ^{\prime}) , \label{G2-gij}
\end{align}
with $i,j=1,2$. We have taken the component $g ^{3} _{\phantom{3} 0}$ to be zero due to the symmetry property of Green's functions. Now we proceed to integrate these equations using the procedure we described in \cref{GF-method}. We obtain
\begin{align}
& g ^{0} _{\phantom{0} i} ( z , z ^{\prime} ) = i \tilde{\theta} \mathfrak{g} ^{(\varepsilon)} (z , 0) \epsilon ^{0j \phantom{k}3} _{\phantom{0j}k} p _{j} g ^{k} _{\phantom{k} i} (0 , z ^{\prime}) , \label{G2-g0i/3} \\ & g ^{i} _{\phantom{i} j} ( z , z ^{\prime} ) =  \eta ^{i} _{\phantom{i}j} \mathfrak{g} ^{(\tilde{\mu})} (z , z ^{\prime}) + i \tilde{\theta} \epsilon ^{0ik3} p _{k} \mathfrak{g} ^{(\tilde{\mu})} (z , 0) g ^{0} _{\phantom{0} j} (0 , z ^{\prime}) - i (1 - \tilde{\mu}) p ^{i} \mathfrak{g} ^{(\tilde{\mu})} (z , 0) g ^{3} _{\phantom{3} j} (0 , z ^{\prime})  . \label{G2-gij/3}
\end{align}
Now we set $z = 0$ in Eq.~(\ref{G2-gij/3}) and then substitute into Eq.~(\ref{G2-g0i/3}), yielding
\begin{align}
& g ^{0} _{\phantom{0} i} ( z , z ^{\prime} ) = i \tilde{\theta} \epsilon ^{0j \phantom{k}3} _{\phantom{0j}i} p _{j} \mathfrak{g} ^{(\varepsilon)} (z , 0) \mathfrak{g} ^{(\tilde{\mu})} (0 , z ^{\prime}) - \tilde{\theta} ^{2} \textbf{p} ^{2} \mathfrak{g} ^{(\tilde{\mu})} (0 , 0) \mathfrak{g} ^{(\varepsilon)} (z , 0) g ^{0} _{\phantom{0} i} (0 , z ^{\prime}) . \label{G2-g0i/4}
\end{align}
Solving for $g ^{0} _{\phantom{0} i} (0 , z ^{\prime})$ by setting $z=0$ in Eq.~(\ref{G2-g0i/4}) and inserting the result back into that equation, we obtain
\begin{align}
g ^{0} _{\phantom{0} i} (z , z ^{\prime}) &= + \frac{i \tilde{\theta} \epsilon ^{0j \phantom{i} 3} _{\phantom{0j} i} p _{j}}{1 + \tilde{\theta} ^{2} \textbf{p} ^{2}  \mathfrak{g} ^{(\varepsilon)} (0 , 0)  \mathfrak{g} ^{(\tilde{\mu})} (0 , 0) } \mathfrak{g} ^{(\varepsilon)} (z , 0) \mathfrak{g} ^{(\tilde{\mu})} (0 , z ^{\prime}) . \label{G2-g0i/FIN}
\end{align}
The remaining components can be obtained by substituting $g ^{0} _{\phantom{0} i} (0 , z ^{\prime})$ in Eq.~(\ref{G2-gij/3}). The result is
\begin{align}
& g ^{i} _{\phantom{i} j} ( z , z ^{\prime} ) =  \eta ^{i} _{\phantom{i}j} \mathfrak{g} ^{(\tilde{\mu})} (z , z ^{\prime}) - \frac{\tilde{\theta} ^{2} \left( \eta ^{i} _{\phantom{i}j} \textbf{p} ^{2} + p ^{i} p _{j} \right) \mathfrak{g} ^{(\varepsilon)} (0 , 0)}{1 + \tilde{\theta} ^{2} \textbf{p} ^{2}  \mathfrak{g} ^{(\varepsilon)} (0 , 0)  \mathfrak{g} ^{(\tilde{\mu})} (0 , 0) } \mathfrak{g} ^{(\tilde{\mu})} (z , 0)  \mathfrak{g} ^{(\tilde{\mu})} (0 , z ^{\prime}) \notag \\ &\hspace{8cm} - (1 - \tilde{\mu}) ^{2} p ^{i} p _{j} \mathfrak{g} ^{(\tilde{\mu})} (0 , 0) \mathfrak{g} ^{(\tilde{\mu})} (z , 0)  \mathfrak{g} (0 , z ^{\prime}) / \tilde{\mu} (z ^{\prime}) , \label{G2-gij/FIN}
\end{align}
where we used the formula $\epsilon ^{0ik3} \epsilon ^{0r \phantom{j} 3} _{\phantom{0r}j} p _{r} p _{k} = \eta ^{i} _{\phantom{i}j} \textbf{p} ^{2} + p ^{i} p _{j}$ and we have substituted the components $g ^{3} _{\phantom{3}j}$ from Eq.~(\ref{G3-gi3-FIN}).

These results can be written in terms of the free space GF $\mathfrak{g} (z,z ^{\prime})$ as follows:
\begin{align}
g ^{0} _{\phantom{0} i} (z , z ^{\prime}) &= + \frac{4 i \tilde{\theta} \epsilon ^{0j \phantom{i} 3} _{\phantom{0j} i} p _{j}}{(\varepsilon + 1)(\frac{1}{\mu} +1) + \tilde{\theta} ^{2} } \mathfrak{g} (z,0) \mathfrak{g} (0,z ^{\prime}) , \label{G2-g0i/FIN2} \\ g ^{i} _{\phantom{i} j} ( z , z ^{\prime} ) &= \eta ^{i} _{\phantom{i} j} \mu (z ^{\prime}) \left[ \mathfrak{g} (z , z ^{\prime}) - 4 \textbf{p} ^{2} \mathfrak{g}(0,0) \mathfrak{g}(z,0) \mathfrak{g}(0,z ^{\prime}) \frac{\mbox{sgn}(z ^{\prime}) (\varepsilon + 1)(\frac{1}{\mu} - 1) + \tilde{\theta}^{2} }{(\varepsilon + 1) (\frac{1}{\mu} + 1) + \tilde{\theta} ^{2}} \right] \notag \\ & \hspace{2cm} - \left[ \frac{2 \tilde{\theta} ^{2} (\frac{1}{\mu} + 1) ^{-1} }{(\varepsilon + 1) (\frac{1}{\mu} + 1) + \tilde{\theta} ^{2}} + \left( \frac{1 - \mu}{1 + \mu} \right) ^{2} \mu (z ^{\prime}) \right] 4 p ^{i} p _{j} \mathfrak{g}(0,0) \mathfrak{g}(z,0) \mathfrak{g}(0,z ^{\prime})  . \label{G2-gij-FIN-App}
\end{align}

\section{GF for a spherical TI} \label{AppSpherical}

In this section we construct the GF for a nonmagnetic spherical topological insulator of radius $a$. Here, the adapted coordinate system is provided by  spherical coordinates. The various components of the GF are the solution of
\begin{equation}
\left[ \mathcal{O} ^{\mu} _{\phantom{\mu} \nu} \right] _{\textbf{x}}  G ^{\nu} _{\; \sigma} (\textbf{x} , \textbf{x} ^{\prime} ) = 4 \pi \eta ^{\mu} _{\phantom{\mu} \sigma} \delta (\textbf{x} - \textbf{x} ^{\prime} ) , \label{GF-Eq-Spherical-App}
\end{equation}
where the differential operator is
\begin{align}
& \left[ \mathcal{O} ^{\mu} _{\phantom{\mu} \nu} \right] _{\textbf{x}} = \left[ \begin{array}{cccc} \mathcal{O} _{r} ^{(\varepsilon)} & \frac{i}{r} \tilde{\theta} \delta (r - a) \hat{\textbf{L}} _{x} & \frac{i}{r} \tilde{\theta} \delta (r - a) \hat{\textbf{L}} _{y} & \frac{i}{r} \tilde{\theta} \delta (r - a) \hat{\textbf{L}} _{z} \\ \frac{i}{r} \tilde{\theta} \delta (r - a) \hat{\textbf{L}} _{x} & \mathcal{O} _{r} ^{(1)} & 0 & 0 \\ \frac{i}{r} \tilde{\theta} \delta (r - a) \hat{\textbf{L}} _{y} & 0 & \mathcal{O} _{r} ^{(1)} & 0 \\ \frac{i}{r} \tilde{\theta} \delta (r - a) \hat{\textbf{L}} _{z} &0 & 0 & \mathcal{O} _{r} ^{(1)} \end{array} \right] . \label{TI-O-operator-Esfericas}
\end{align}
Here $\hat{\textbf{L} _{k}}$ are the components of the angular momentum operator, and
\begin{align}
\mathcal{O} _{r} ^{(\varepsilon)} =  - \varepsilon (r) \nabla ^{2} - \frac{\partial \varepsilon (r)}{\partial r} \frac{\partial}{\partial r} ,
\end{align}
and $\mathcal{O} _{r} ^{(1)} = - \nabla ^{2}$. Since the completeness relation for the spherical harmonics is 
\begin{align}
\delta (\cos \vartheta - \cos \vartheta ^{\prime}) \delta (\varphi - \varphi ^{\prime}) = \sum _{l=0} ^{\infty} \sum _{m = -l} ^{+l} Y _{lm} (\vartheta , \varphi) Y _{lm} ^{\ast} (\vartheta ^{\prime} , \varphi ^{\prime}) ,
\end{align}
we look for a solution of the form
\begin{equation}
G ^{\mu} _{\phantom{\mu} \nu} (\textbf{x} , \textbf{x} ^{\prime} ) = 4 \pi \sum _{l=0} ^{\infty} \sum _{l ^{\prime} =0} ^{\infty} \sum _{m=-l} ^{+l} \sum _{m ^{\prime}=-l} ^{+l} g ^{\mu} _{l l ^{\prime} m m ^{\prime} , \nu} (r , r ^{\prime} ) Y _{lm} (\vartheta , \varphi) Y _{l ^{\prime} m ^{\prime}} ^{\ast} (\vartheta ^{\prime} , \varphi ^{\prime}) ,  \label{TI-GF-Esfericas}
\end{equation}
where $g ^{\mu} _{l l ^{\prime }m m ^{\prime} , \nu} (r , r ^{\prime} )$ is the reduced GF analogous to $g ^{\mu} _{\phantom{\mu} \nu} (z,z ^{\prime})$ in the case of a planar TI. The operator (\ref{TI-O-operator-Esfericas}) commutes with $\hat{\textbf{L}} ^{2}$ in such a way that
\begin{align}
g ^{\mu} _{l l ^{\prime} m m ^{\prime} , \nu} (r , r ^{\prime} ) = \delta _{ll ^{\prime}} g ^{\mu} _{l m m ^{\prime} , \nu} (r , r ^{\prime} ) .
\end{align}
In the following, we focus on determining the various components of the GF matrix in Eq. (\ref{GF-Eq-Spherical-App}). The method we shall employ is similar to that used for solving the planar case, but the required mathematical techniques are more subtle because the dependence upon the angular momentum operator.

Substituting Eq. (\ref{TI-GF-Esfericas}) into Eq. (\ref{GF-Eq-Spherical-App}), and using the properties of the spherical harmonics we find that the reduced GF satisfies the differential equation
\begin{align}
\sum _{m ^{\prime \prime}=-l} ^{+l} \hat{R} ^{\mu} _{l m m ^{\prime \prime} , \nu}  g ^{\nu} _{l m ^{\prime \prime} m ^{\prime} , \sigma} (r , r ^{\prime} ) = \eta ^{\mu} _{\phantom{\mu}  \sigma} \delta _{m m ^{\prime}} \frac{\delta (r - r ^{\prime})}{r ^{2}} , \label{TI-GF/Matrix-Eq-Esfericas3}
\end{align}
where we defined the radial operator
\begin{equation}
\hat{R} ^{\mu} _{ l m m ^{\prime \prime} , \nu} = \langle l m \vert \left[ \mathcal{O} ^{\mu} _{\phantom{\mu} \nu} \right] _{\textbf{x}} \vert l m ^{\prime \prime} \rangle = \int _{\Omega} Y _{lm} ^{\ast} (\vartheta , \varphi) \left[ \mathcal{O} ^{\mu} _{\phantom{\mu} \nu} \right] _{\textbf{x}} Y _{lm ^{\prime \prime}} (\vartheta , \varphi) d \Omega , \label{R-Operator}
\end{equation}
which can be written in the matrix form
\begin{align}
\hat{R} ^{\mu} _{ l m m ^{\prime \prime} , \nu} = \left[ \begin{array}{cccc} \delta _{m m ^{\prime \prime}} \hat{\mathcal{O}} _{r} ^{(\varepsilon)}  & \frac{i}{r} \tilde{\theta} \delta (r - a) \langle \hat{\textbf{L}} _{x} \rangle _{m m ^{\prime \prime}} & \frac{i}{r} \tilde{\theta} \delta (r - a) \langle \hat{\textbf{L}} _{y} \rangle _{m m ^{\prime \prime}} & \frac{i}{r} \tilde{\theta} \delta (r - a) \langle \hat{\textbf{L}} _{z} \rangle _{m m ^{\prime \prime}} \\ \frac{i}{r} \tilde{\theta} \delta (r - a) \langle \hat{\textbf{L}} _{x} \rangle _{m m ^{\prime \prime}} & \delta _{m m ^{\prime \prime}} \hat{\mathcal{O}} _{r} ^{(1)} & 0 & 0 \\ \frac{i}{r} \tilde{\theta} \delta (r - a) \langle \hat{\textbf{L}} _{y} \rangle _{m m ^{\prime \prime}} & 0 & \delta _{m m ^{\prime \prime}} \hat{\mathcal{O}} _{r} ^{(1)} & 0 \\ \frac{i}{r} \tilde{\theta} \delta (r - a) \langle \hat{\textbf{L}} _{z} \rangle _{m m ^{\prime \prime}} & 0 & 0 & \delta _{m m ^{\prime \prime}} \hat{\mathcal{O}} _{r} ^{(1)} \end{array} \right] . \label{R-Operator2}
\end{align}
The notation here is as follows: ($\hat{\textbf{L}}_i=-\hat{\textbf{L}}^i, i=x,y,z$)
\begin{align}
\langle \hat{\textbf{L}} _{i} \rangle _{m m ^{\prime \prime}} &= \langle l m  \vert \hat{\textbf{L}} _{i} \vert lm^{\prime \prime} \rangle = \int _{\Omega} Y _{lm} ^{\ast} (\vartheta , \varphi) \hat{\textbf{L}} _{i} Y _{lm ^{\prime \prime}} (\vartheta , \varphi) d \Omega ,  \label{CompMomAngular} \\ \hat{\mathcal{O}} _{r} ^{(\varepsilon)} &= \varepsilon (r) \hat{\mathcal{O}} _{r} ^{(1)} - \frac{\partial \varepsilon (r)}{\partial r} \frac{\partial}{\partial r} , \\ \hat{\mathcal{O}} _{r} ^{(1)}  &= \frac{l (l+1)}{r ^{2}} - \frac{1}{r ^{2}} \frac{\partial}{\partial r} \left( r ^{2} \frac{\partial }{\partial r} \right) .
\end{align}
The resulting equation for the components of the reduced GF (\ref{TI-GF/Matrix-Eq-Esfericas3}) can be integrated using the reduced Green's functions associated with the operators $\hat{\mathcal{O}} _{r} ^{(1)}$ and $\hat{\mathcal{O}} _{r} ^{(\varepsilon)}$ previously defined, that solve
\begin{align}
\hat{\mathcal{O}} _{r} ^{(1)} \mathfrak{g} ^{(1)} _{l} (r,r ^{\prime}) = \frac{\delta (r - r ^{\prime})}{r ^{2}} , \\ \hat{\mathcal{O}} _{r} ^{(\varepsilon)} \mathfrak{g} ^{(\varepsilon)} _{l} (r,r ^{\prime}) = \frac{\delta (r - r ^{\prime})}{r ^{2}} ,
\end{align}
satisfying the standard boundary conditions at infinity. In the former we have the free space solution
\begin{align}
\label{g1l_freeSpher}
\mathfrak{g} ^{(1)} _{l} (r,r ^{\prime}) = \frac{r _{<} ^{l}}{r _{>} ^{l+1}} \frac{1}{2l + 1}
\end{align}
where $r _{>}$ ($r _{<}$) is the greater (lesser) or $r$ and $r ^{\prime}$; and for the second case we have the reduced GF for a dielectric sphere \cite{Schwinger}.
\begin{align}
r ^{\prime} > a, r > a \quad &: \quad \mathfrak{g} ^{(\varepsilon)} _{l} (r,r ^{\prime}) = \frac{1}{r _{>} ^{l+1}} \frac{1}{2l+1} \left[ r _{<} ^{l} - \frac{(\varepsilon - 1) l}{(\varepsilon + 1) l + 1} \frac{a ^{2l+1}}{r _{<} ^{l+1}} \right]  , \label{gel_1}\\ r ^{\prime} > a, r < a \quad &: \quad \mathfrak{g} ^{(\varepsilon)} _{l} (r,r ^{\prime}) = \frac{1}{(\varepsilon + 1) l + 1} \frac{r ^{l}}{r ^{\prime \; l+1}} , \label{gel_2}\\ r ^{\prime} < a, r < a \quad &: \quad \mathfrak{g} ^{(\varepsilon)} _{l} (r,r ^{\prime}) = \frac{r _{<} ^{l} }{(2l+1) \varepsilon} \left[\frac{1}{r _{>} ^{l+1}} + \frac{(\varepsilon - 1) (l +1)}{(\varepsilon + 1) l + 1} \frac{r _{>} ^{l}}{a ^{2l+1}} \right]  , \label{gel_3} \\ r ^{\prime} < a, r > a \quad &: \quad \mathfrak{g} ^{(\varepsilon)} _{l} (r,r ^{\prime}) = \frac{1}{(\varepsilon + 1) l + 1} \frac{r ^{\prime \; l}}{r ^{l+1}} . \label{gel_4}
\end{align}
Now we proceed to integrate Eq.~(\ref{TI-GF/Matrix-Eq-Esfericas3}). We observe that the corresponding equations for the reduced GF fall into two groups. The first is defined by $\sigma = 0$ and the rest by $\sigma = i$ in Eq.~(\ref{TI-GF/Matrix-Eq-Esfericas3}). The four equations for the first group are 
\begin{align}
& \hat{\mathcal{O}} ^{(\varepsilon)} _{r} g ^{0} _{lmm^{\prime},0}- \frac{i}{r} \tilde{\theta} \delta (r-a) \sum _{m ^{\prime \prime} = -l} ^{+l} \langle \hat{\textbf{L}} _{i} \rangle _{m m ^{\prime \prime}} g ^{i} _{lm ^{\prime \prime} m^{\prime},0} = \delta _{mm ^{\prime}} \frac{\delta (r - r ^{\prime})}{r ^{2}} , \\ & \hat{\mathcal{O}} ^{(1)} _{r} g ^{i} _{lmm^{\prime},0} + \frac{i}{r} \tilde{\theta} \delta (r-a) \eta ^{ij} \sum _{m ^{\prime \prime} = -l} ^{+l} \langle \hat{\textbf{L}} _{j} \rangle _{m m ^{\prime \prime}} g ^{0} _{lm ^{\prime \prime} m^{\prime},0} = 0 ,
\end{align}
which can be integrated to yield
\begin{align}
g ^{0} _{lmm^{\prime},0} (r,r ^{\prime}) &= \delta _{mm ^{\prime}} \mathfrak{g} ^{(\varepsilon)} _{l} (r,r ^{\prime}) + i a \tilde{\theta} \mathfrak{g} ^{(\varepsilon)} _{l} (r,a) \sum _{m ^{\prime \prime} = -l} ^{+l} \langle \hat{\textbf{L}} _{i} \rangle _{m m ^{\prime \prime}} g ^{i} _{lm ^{\prime \prime} m^{\prime},0} (a,r ^{\prime}) , \label{g00-esf} \\ g ^{i} _{lmm^{\prime},0} (r,r ^{\prime}) &= - i a \tilde{\theta}  \mathfrak{g} ^{(1)} _{l} (r,a) \sum _{m ^{\prime \prime} = -l} ^{+l} \langle \hat{\textbf{L}}^{i} \rangle _{m m ^{\prime \prime}} g ^{0} _{lm ^{\prime \prime} m^{\prime},0} (a,r ^{\prime}) . \label{gi0-esf}
\end{align}
Now, we set $r=a$ in Eq.~(\ref{gi0-esf}) and then substitute into Eq.~(\ref{g00-esf}) yielding
\begin{align}
g ^{0} _{lmm^{\prime},0} (r,r ^{\prime}) &= \delta _{mm ^{\prime}} \mathfrak{g} ^{(\varepsilon)} _{l} (r,r ^{\prime}) - a ^{2} \tilde{\theta} ^{2} l (l+1) \mathfrak{g} ^{(1)} _{l} (a,a) \mathfrak{g} ^{(\varepsilon)} _{l} (r,a) g ^{0} _{lmm^{\prime},0} (a,r ^{\prime}) , \label{g00-esf-2} 
\end{align}
where we have used the result $\eta ^{ij} \sum _{m ^{\prime}} \langle \hat{\textbf{L}} _{i} \rangle _{m m ^{\prime}} \langle \hat{\textbf{L}} _{j} \rangle _{m ^{\prime} m ^{\prime \prime}} = - l (l+1) \delta _{mm ^{\prime \prime}} $. Solving for $g ^{0} _{lmm^{\prime},0} (a,r ^{\prime})$ by setting $r=a$ in Eq.~(\ref{g00-esf-2}) and inserting the result back into that equation, we obtain
\begin{align}
g ^{0} _{lmm^{\prime},0} (r,r ^{\prime}) &= \delta _{mm ^{\prime}} \left[ \mathfrak{g} ^{(\varepsilon)} _{l} (r,r ^{\prime}) - a ^{2} \tilde{\theta} ^{2} l (l+1) \mathfrak{g} ^{(1)} _{l} (a,a) S _{l} ^{(\varepsilon , \varepsilon)} (r,r ^{\prime}) \right] , \label{g00-esf-FIN} 
\end{align}
where the function $S _{l} ^{(\varepsilon , \varepsilon)} (r,r ^{\prime}) $ was defined in Eq.~(\ref{S-function}). The remaining components can be computed directly by substituting $g ^{0} _{lmm^{\prime},0} (a,r ^{\prime})$ in Eq.~(\ref{gi0-esf}). The result is
\begin{align}
g ^{i} _{lmm^{\prime},0} (r,r ^{\prime}) &= - i a \tilde{\theta} \langle \hat{\textbf{L}}^{i} \rangle _{m m ^{\prime}} S _{l} ^{(1 , \varepsilon)} (r,r ^{\prime}) . \label{gi0-esf-FIN}
\end{align}
The second group of equations defined by $\sigma = i$ in Eq.~(\ref{TI-GF/Matrix-Eq-Esfericas3}) is
\begin{align}
& \hat{\mathcal{O}} ^{(\varepsilon)} _{r} g ^{0} _{lmm^{\prime},i}- \frac{i}{r} \tilde{\theta} \delta (r-a) \sum _{m ^{\prime \prime} = -l} ^{+l} \langle \hat{\textbf{L}} _{j} \rangle _{m m ^{\prime \prime}} g ^{j} _{lm ^{\prime \prime} m^{\prime},i} = 0, \\ & \hat{\mathcal{O}} ^{(1)} _{r} g ^{j} _{lmm^{\prime},i} + \frac{i}{r} \tilde{\theta} \delta (r-a) \eta ^{jk} \sum _{m ^{\prime \prime} = -l} ^{+l} \langle \hat{\textbf{L}} _{k} \rangle _{m m ^{\prime \prime}} g ^{0} _{lm ^{\prime \prime} m^{\prime},i} = \delta _{mm ^{\prime}} \frac{\delta (r - r ^{\prime})}{r ^{2}}  .
\end{align}
Integrating these equation we obtain
\begin{align}
g ^{0} _{lmm^{\prime},i} (r,r ^{\prime}) &= i a \tilde{\theta} \mathfrak{g} ^{(\varepsilon)} _{l} (r,a) \sum _{m ^{\prime \prime} = -l} ^{+l} \langle \hat{\textbf{L}} _{j} \rangle _{m m ^{\prime \prime}} g ^{j} _{lm ^{\prime \prime} m^{\prime},i} (a,r ^{\prime}) , \label{g0i-esf} \\ g ^{j} _{lmm^{\prime},i} (r,r ^{\prime}) &= \delta _{mm ^{\prime}} \eta ^{j} _{\phantom{j}i} \mathfrak{g} ^{(1)} _{l} (r,r ^{\prime}) - i a \tilde{\theta} \mathfrak{g} ^{(1)} _{l} (r,a) \sum _{m ^{\prime \prime} = -l}^{+l} \langle \hat{\textbf{L}}^{j} \rangle _{m m ^{\prime \prime}} g ^{0} _{lm ^{\prime \prime} m^{\prime},i} (r , r ^{\prime})  . \label{gji-esf}
\end{align}
Now we solve in the same way as for the previous group of equations. Setting $r = a$ in Eq.~(\ref{gji-esf}) and then substituting into Eq.~ (\ref{g0i-esf}) yields
\begin{align}
g ^{0} _{lmm^{\prime},i} (r,r ^{\prime}) &= i a \tilde{\theta}  \langle \hat{\textbf{L}} _{i} \rangle _{m m ^{\prime}} \mathfrak{g} ^{(\varepsilon)} _{l} (r,a) \mathfrak{g} ^{(1)} _{l} (a,r ^{\prime}) - a ^{2} \tilde{\theta} ^{2} l (l+1) \mathfrak{g} ^{(1)} _{l} (a,a) \mathfrak{g} ^{(\varepsilon)} _{l} (r,a) g ^{0} _{lmm^{\prime},i} (a,r ^{\prime}) . \label{g0i-esf-2}
\end{align}
Solving for $g ^{0} _{lmm^{\prime},i} (a,r ^{\prime})$ by setting $r=a$ in Eq.~(\ref{g0i-esf-2}) and inserting the result back in this equation, we obtain
\begin{align}
g ^{0} _{lmm^{\prime},i} (r,r ^{\prime}) &= i a \tilde{\theta}  \langle \hat{\textbf{L}} _{i} \rangle _{m m ^{\prime}} S _{l} ^{(\varepsilon , 1)} (r,r ^{\prime}) , \label{g0i-esf-FIN}
\end{align}
where the function $S _{l} ^{(\varepsilon , 1)} (r,r ^{\prime})$ was defined in Eq.~(\ref{S-function}). The remaining components can be computed similarly. The substitution of $g ^{0} _{lmm^{\prime},i} (a,r ^{\prime})$ in Eq.~(\ref{gji-esf}) yields
\begin{align}
g ^{i} _{l m m^{\prime} , j} (r , r ^{\prime} ) &= \eta ^{i} _{\phantom{i} j} \delta _{mm ^{\prime}} \mathfrak{g} ^{(1)} _{l} (r , r ^{\prime} ) + a ^{2} \tilde{\theta} ^{2} \langle \hat{\textbf{L}} ^{i} \hat{\textbf{L}} _{j} \rangle _{mm ^{\prime}} \mathfrak{g} ^{(\varepsilon)} _{l} (a , a ) S _{l} ^{( 1,1)} (r , r ^{\prime}) .
\end{align}
These results establish Eqs.~(\ref{g00-esf-F})-(\ref{gij-esf-F}).


\begin{thebibliography}{99}

\bibitem{Anderson} P. W. Anderson, \textit{Basic Notions of Condensed Matter Physics} (Westview Press, Boulder, CO, 1997).

\bibitem{Qi-ReviewTI} X. L. Qi and S. C Zhan, Rev. Mod. Phys. \textbf{83}, 1057 (2011).

\bibitem{Hasan-ReviewTI} M. Z. Hasan and C. L. Kane, Rev. Mod. Phys. \textbf{82}, 3045 (2010).

\bibitem{Wang} Z. Wang, X. L. Qi and S. C. Zhang, Phys. Rev. Lett. \textbf{105}, 256803 (2010).

\bibitem{Maciejko-PRL} J. Maciejko, X. L. Qi, A. Karch and S. C. Zhang, Phys. Rev. Lett. \textbf{105}, 246809 (2010).

\bibitem{Zhou} X. Z. Zhou, J. Zhang, X. Ling, S. Chen, H. Luo and S. Wen, Phys. Rev. A \textbf{88}, 053840 (2013).

\bibitem{Qi-PRB} X. L. Qi, T. L. Hughes and S. C Zhang, Phys. Rev. B \textbf{78}, 195424 (2008).

\bibitem{Essin} A. M. Essin, J. E. Moore and D. Vanderbilt, Phys. Rev. Lett. \textbf{102}, 146805 (2009).

\bibitem{Qi-Science} X. L. Qi, R. Li, J. Zang and S. C. Zhang, Science \textbf{323}, 1184 (2009).

\bibitem{Rosenberg} G. Rosenberg and M. Franz, Phys. Rev. B \textbf{82}, 035105 (2010).

\bibitem{Hehl} Y. N. Obukhov and F. W. Hehl, Phys. Lett. A \textbf{341}, 357 (2005).

\bibitem{Maciejko} J. Maciejko, X. L. Qi, H. D. Drew and S. C. Zhang, Phys. Rev. Lett. \textbf{105}, 166803 (2010).

\bibitem{Huerta} L.~Huerta and J.~Zanelli, Phys.\ Rev.\ D {\bf 85}, 085024 (2012); L. Huerta, Phys. Rev. D \textbf{90}, 105026 (2014).

\bibitem{Grushin-PRL} A. G. Grushin and A. Cortijo, Phys. Rev. Lett. \textbf{106}, 020403 (2011); A. G. Grushin, P. Rodriguez-Lopez and A. Cortijo, Phys. Rev. B. \textbf{84}, 045119 (2011).

\bibitem{Rodriguez-PRL} P. Rodriguez-Lopez and A. G. Grushin, Phys. Rev. Lett. \textbf{112}, 056804 (2014).

\bibitem{MCU1} A. Mart\'{i}n-Ruiz, M. Cambiaso and L. F. Urrutia, Phys. Rev. D \textbf{92}, 125015 (2015); Phys. Rev. D \textbf{93}, 045022 (2016).

\bibitem{MCU3} A. Mart\'{i}n-Ruiz, M. Cambiaso and L. F. Urrutia, Eur. Phys. Lett. \textbf{113}, 60005 (2016).

\bibitem{Schwinger} J. Schwinger, L. DeRaad, K. Milton and W. Tsai, ``Classical Electrodynamics", Advanced Book Program, (Perseus Books 1998).

\bibitem{Karch} A. Karch, Phys. Rev. Lett. \textbf{103}, 171601 (2009).

\bibitem{Dzyaloshinskii2} I. Dzyaloshinskii,  Sov. Phys. JETP,  \textbf{36}, 1797-1805 (1959).

\bibitem{Casimir}%
D. Kupiszewska and J. Mostowski, Phys. Rev. A 41, 4636 (1990);
T.G. Philbin, C. Xiong, U. Leonhardt, Annals of Physics 325 (2010) 579–595;
S.~J.~Rahi, T.~Emig and R.~L.~Jaffe,
Lect.\ Notes Phys.\  {\bf 834}, 129 (2011);
S.~J.~Rahi, T.~Emig, N.~Graham, R.~L.~Jaffe and M.~Kardar,
Phys.\ Rev.\ D {\bf 80}, 085021 (2009);
W. M. R. Simpson, S. A. R. Horsley, U. Leonhardt,   Phys. Rev. A \textbf{87}, 043806 (2013);
S. A. R. Horsley, W. M. R. Simpson,    Phys. Rev. A \textbf{88}, 013833 (2013).

\bibitem{Crosse:2015loa} 
  J.~A.~Crosse, S.~Fuchs and S.~Y.~Buhmann,
  Phys.\ Rev.\ A {\bf 92}, no. 6, 063831 (2015).

\bibitem{LL_EM}
Landau, L. D., Lifshitz, E. M., and Pitaevskii, L. P. (1984). Electrodynamics of
Continuous Media. Pergamon, Oxford.

\bibitem{Dzyaloshinskii} I. Dzyaloshinskii Zh. Exp. Teor. Fiz. \textbf{37}, 881 (1960).

\end{thebibliography}
\end{document}